\PassOptionsToPackage{table}{xcolor}
\documentclass[11pt]{article}

\usepackage[final]{acl}

\usepackage{times}
\usepackage{latexsym}
\usepackage[T1]{fontenc}
\usepackage[utf8]{inputenc}
\usepackage{microtype}
\usepackage{inconsolata}
\usepackage{enumitem}

\usepackage{adjustbox}
\usepackage{booktabs}
\usepackage{graphicx}
\usepackage{xcolor}
\usepackage{array}
\usepackage{tabularx}

\usepackage{amsmath}
\usepackage{amssymb}
\usepackage{pifont}

\newcommand{\cmark}{\ding{51}}
\newcommand{\xmark}{\textcolor{gray!45}{\ding{55}}}
\newcolumntype{L}[1]{>{\raggedright\arraybackslash}p{#1}}
\newcolumntype{C}[1]{>{\centering\arraybackslash}p{#1}}
\newcolumntype{R}[1]{>{\raggedleft\arraybackslash}p{#1}}

\usepackage[most]{tcolorbox}
\usepackage{listings}
\usepackage{multirow}
\definecolor{stdGray}{HTML}{52514E}
\definecolor{colImplicit}{HTML}{0084FF}    
\definecolor{colCumulative}{HTML}{FF4A4F}  
\definecolor{colUnmet}{HTML}{C363FF}       
\colorlet{colStandard}{stdGray}            
\definecolor{okGreen}{HTML}{2F8A55}
\definecolor{badRed}{HTML}{CF5D55}
\lstdefinestyle{promptstyle}{
  basicstyle=\ttfamily\fontsize{6.8pt}{8.2pt}\selectfont,
  breaklines=true, breakatwhitespace=true, columns=fullflexible,
  keepspaces=true, showstringspaces=false, upquote=true,
  breakindent=0pt, postbreak={}, tabsize=2, aboveskip=0pt, belowskip=0pt,
  inputencoding=utf8, extendedchars=true, literate={✓}{{\ding{51}}}1}
\newcommand{\promptfile}[3][clBlue]{%
  \tcbinputlisting{enhanced, breakable, listing only, listing file={#3},
    listing options={style=promptstyle},
    colframe=#1!75!black, colback=#1!3, colbacktitle=#1!75!black,
    coltitle=white, fonttitle=\small\bfseries, title={#2},
    boxrule=0.6pt, arc=2pt, left=4pt, right=4pt, top=3pt, bottom=3pt}}
\newtcolorbox{examplebox}[2][clBlue]{enhanced, breakable,
  colframe=#1!75!black, colback=#1!3, colbacktitle=#1!75!black,
  coltitle=white, fonttitle=\small\bfseries, title={#2},
  fontupper=\footnotesize, boxrule=0.6pt, arc=2pt,
  left=4pt, right=4pt, top=3pt, bottom=3pt}
\newcommand{\ctype}[1]{\textsc{#1}}
\newcommand{\clevel}[1]{\textit{#1}}
\newcommand{\evid}[1]{\leavevmode\textcolor{black!60}{``#1''}}

\definecolor{retrievalrow}{RGB}{224,240,248}
\definecolor{corpusagentrow}{RGB}{235,228,247}
\definecolor{webagentrow}{RGB}{229,243,233}

\usepackage{tikz}\usetikzlibrary{fadings}
\definecolor{clBlue}{HTML}{30B0F0}
\definecolor{clPink}{HTML}{F07080}
\definecolor{clPurple}{HTML}{C050F0}
\pgfdeclarehorizontalshading{complitgrad}{100bp}{
  color(0bp)=(clBlue);      
  color(39bp)=(clBlue);     
  color(39bp)=(clPink);     
  color(60bp)=(clPink);     
  color(60bp)=(clPurple);   
  color(100bp)=(clPurple)   
}

\begin{tikzfadingfrompicture}[name=complitfadeT]
  \node [text=transparent!0, font=\Large\bfseries\scshape] {CompLit};
\end{tikzfadingfrompicture}
\begin{tikzfadingfrompicture}[name=complitfadeR]
  \node [text=transparent!0, font=\normalsize\scshape] {CompLit};
\end{tikzfadingfrompicture}

\makeatletter
\newcount\cl@n
\newcommand{\CompLitPat}[3]{
  \leavevmode\begingroup
  \setbox0=\hbox{#2CompLit}%
  \pgfmathsetmacro{\clW}{\wd0/1.00375}
  \pgfmathsetmacro{\clHt}{\ht0/1.00375}\pgfmathsetmacro{\clDp}{\dp0/1.00375}%
  \pgfmathsetmacro{\clH}{\clHt+\clDp}%
  \pgfmathsetmacro{\clCx}{0.5*\clW}\pgfmathsetmacro{\clCy}{0.5*\clH}%
  \pgfmathsetmacro{\clSx}{1.41421*(\clW*abs(cos(#1))+\clH*abs(sin(#1)))/100}
  \pgfmathsetmacro{\clA}{50*\clSx}%
  \pgfmathsetmacro{\clPax}{\clCx-\clA*cos(#1)}\pgfmathsetmacro{\clPay}{\clCy-\clA*sin(#1)}%
  \pgfmathsetmacro{\clPbx}{\clCx+\clA*cos(#1)}\pgfmathsetmacro{\clPby}{\clCy+\clA*sin(#1)}%
  \immediate\pdfobj{<< /PatternType 2 /Shading << /ShadingType 2 /ColorSpace /DeviceRGB
    /Coords [\clPax\space \clPay\space \clPbx\space \clPby] /Extend [true true]
    /Function << /FunctionType 3 /Domain [0 1] /Bounds [0.39 0.60] /Encode [0 1 0 1 0 1]
      /Functions [ << /FunctionType 2 /Domain [0 1] /N 1 /C0 [0.188 0.690 0.941] /C1 [0.188 0.690 0.941] >>
                   << /FunctionType 2 /Domain [0 1] /N 1 /C0 [0.941 0.439 0.502] /C1 [0.941 0.439 0.502] >>
                   << /FunctionType 2 /Domain [0 1] /N 1 /C0 [0.753 0.314 0.941] /C1 [0.753 0.314 0.941] >> ] >> >> >>}%
  \edef\cl@pat{\the\pdflastobj}%
  \global\advance\cl@n by 1
  \pgfmathsetmacro{\clSw}{#3}%
  \setbox2=\hbox{\pdfliteral page{/Pattern cs /CLpat\the\cl@n\space scn /Pattern CS /CLpat\the\cl@n\space SCN 1 j 1 J \clSw\space w \ifdim\clSw pt>0pt 2\else 0\fi\space Tr}\copy0}%
  \immediate\pdfxform resources{/Pattern << /CLpat\the\cl@n\space \cl@pat\space 0 R >>}2%
  \pdfrefxform\pdflastxform
  \endgroup}
\makeatother
\newcommand{\CompLitTitle}[1]{\CompLitPat{#1}{\Large\bfseries\scshape}{0}}
\newcommand{\CompLitRegular}[1]{\CompLitPat{#1}{\normalsize\scshape}{0.18}}

\newcommand{\bestscore}[1]{\textbf{#1}}

\usepackage{xspace}
\newcommand{\setname}[1]{\textsc{#1}}
\newcommand{\Std}{\setname{Standard}\xspace}
\newcommand{\RI}{\setname{Implicit}\xspace}
\newcommand{\CI}{\setname{Cumulative}\xspace}
\newcommand{\NM}{\setname{Unmet}\xspace}

\title{
\texorpdfstring{\CompLitTitle{310}}{CompLit}: Scientific Literature Search Benchmarks Must Cover
\\ Implicit, Cumulative, and Unmet Needs
}

\author{
  \textbf{Dayoon Ko\textsuperscript{1}}\thanks{Work was done during internship at LG AI Research.}
  \textbf{Jihyuk Kim\textsuperscript{2}}
  \textbf{Soyeong Jeong\textsuperscript{3}}
  \textbf{Young-Jun Lee\textsuperscript{4}}
  \textbf{Dahyun Lee\textsuperscript{2}}
\\
  \textbf{Juyeon Kim\textsuperscript{2}}
  \textbf{Gunhee Kim\textsuperscript{1}}
  \textbf{Moontae Lee\textsuperscript{2,5}}
  \textbf{Kyungjae Lee\textsuperscript{2,6}}\thanks{Corresponding author.}
\\
\vspace{-3mm}
\\\small{
  \textsuperscript{1}Seoul National University
  \textsuperscript{2}LG AI Research
  \textsuperscript{3}KAIST}
  \\\small{
  \textsuperscript{4}University of Minnesota
  \textsuperscript{5}University of Illinois Chicago
  \textsuperscript{6}University of Seoul
  }
\\
  \small{
  \texttt{
  \small{\url{https://dayoon-ko.github.io/CompLit/}}
  }
  }
}

\begin{document}
\maketitle

\begin{abstract}
How much can we trust an agent to find the paper we need? As researchers, we search the literature from what we have read toward what we have not. We thus describe what we want in our own terms, which a paper may show only implicitly; our requirements grow as we read; and what we seek may lie in an empty region no paper yet covers. We introduce \textbf{\textsc{CompLit}}, a literature search benchmark spanning all eight arXiv domains, with one diagnostic setting for each of these needs. The strongest agents, Codex (GPT-5.6-Sol) and Claude Code (Opus 4.8), answer 86--87\% of \Std queries, yet fall to 66--68\% when requirements are implicit (\RI), lose another 6 and 15pp, respectively, as they accumulate (\CI), and abstain on at most 53\% of the queries no paper satisfies (\NM). No agent leads in every domain, and all are less accurate on papers from before 2019 or after 2023. Tracing their search, we find that agents often reach the right paper, or the closest one, yet misjudge it. Doing well on \textsc{CompLit} is a step toward agents that we can trust to find the paper we need, or to tell us that none exists.

\end{abstract}


\section{Introduction}

Researchers increasingly ask agents such as OpenAI Deep Research, Codex, or Claude Code to find papers they need \citep{openai2025deepresearch,
anthropic2025claudecode}. The paper an agent returns shapes what they build on, what they cite, and whether they consider their ideas new 
\citep{greenberg2009citation, pavlovic2021accurate}. 

Despite these stakes, most literature search benchmarks (Table~\ref{tab:benchmark-comparison}) evaluate agents from the paper's side rather than the researcher's. They build each query from a target paper, so it reuses the paper's terms, the whole request comes in a single query, and a matching paper always exists. Such queries are thus bound to the target paper: its wording, a single request, and an existing answer.

\begin{figure}[t!]
    \vspace{-5mm}
    \centering
    \includegraphics[width=1\linewidth]{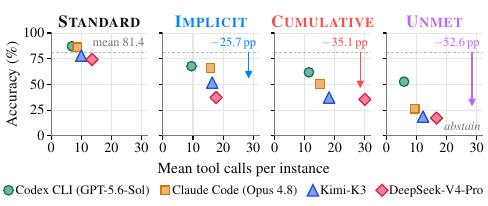} 
    \vspace{-7mm}
    \caption{
    \textbf{Frontier agents on \textsc{CompLit}.} All four agents score well on \Std (mean 81.4\%) but drop on every diagnostic setting (arrows: drop in mean). On \NM, accuracy is the rate of correct abstention. 
    }
    \label{fig:intro_teaser}
    \vspace{-5mm}
\end{figure}

\begin{figure*}[t!]
    \centering
    \includegraphics[width=1\linewidth]{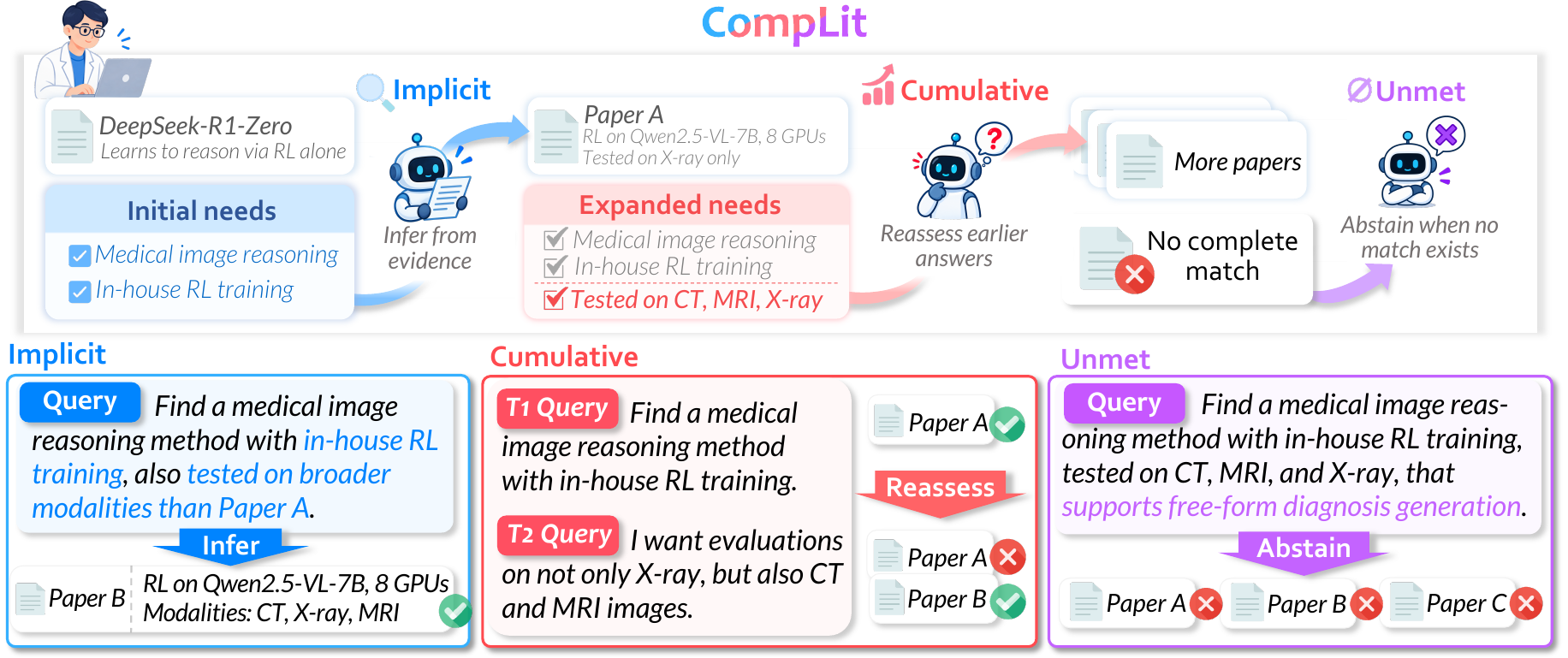} 
    \vspace{-6mm}
    \caption{\textbf{Motivation and overview of \textsc{CompLit}.} \textbf{Top:} as a researcher reads retrieved papers, their need proves \textit{implicit}, \textit{cumulative}, and \textit{unmet}: the required property is rarely stated outright, each new requirement rules out earlier candidates, and the full set may match no paper. \textbf{Bottom:} our three diagnostic settings test whether (i) an agent \textit{infers} such a property from paper evidence, (ii) \textit{reassesses} earlier answers as requirements accumulate, and (iii) \textit{abstains} when no paper meets the need.}
    \label{fig:intro}
    \vspace{-4mm}
\end{figure*}

Researchers, however, search outward, from what they have explored toward what they have not \citep{kuhlthau1991inside}. Because they frame requirements from where they stand, the paper they seek may state them only \textit{implicitly} \citep{dasigi-etal-2021-dataset}; as they read on, requirements \textit{accumulate} \citep{vakkari2001theory}; and as they seek what no one has explored, their need may be \textit{unmet} \citep{asai2021challenges}.

In Figure~\ref{fig:intro}, a medical imaging researcher cannot send patient scans outside the hospital. After reading how DeepSeek-R1-Zero learns to reason through reinforcement learning alone \citep{guo2025deepseek}, they look for a method that does the same in-house. A matching paper often describes how it trains without explicitly stating that it does so in-house (\textit{implicit}). Reading that such a paper tests only on X-ray, the researcher also asks for CT and MRI, ruling it out (\textit{cumulative}). If no paper meets them all, the direction may still be open (\textit{unmet}).


We introduce \textsc{CompLit}, a benchmark for compositional scientific literature search that evaluates agents in these situations. Alongside a \Std reference setting close to existing benchmarks, it has one diagnostic setting per need, each requiring a distinct ability. \RI requires an agent to \textit{infer} whether a paper meets a requirement from how the paper describes its work. \CI requires an agent to \textit{reassess} its earlier answers and move beyond them as requirements accumulate over turns. \NM requires an agent to \textit{abstain} when no paper satisfies every requirement.

We compose each query from several requirements taken from survey comparison tables, and collect 870 single-turn queries and 220 conversations, all manually verified, across all eight arXiv domains. Beyond final accuracy and its breakdown by domain and publication year, \textsc{CompLit} traces agents' search dynamics over tool calls or turns for each setting, pinpointing where they go wrong.


The strongest agents already reach 86--87\% accuracy on \Std (Figure~\ref{fig:intro_teaser}), where, as in existing benchmarks, most requirements are stated outright, all arrive at once, and a match always exists. 
Even there, average accuracy ranges from 74\% in physics to 94\% in finance and economics, and falls from 88\% on papers from 2019--23 to 75\% on older and 80\% on more recent ones (Table~\ref{tab:standard-breakdown}). 
Beyond these gaps, each diagnostic setting reveals a different way they fall short.

On \RI, agents struggle to \textit{infer} whether a paper meets a requirement from indirect evidence. The gold paper is retrieved less often, and even when it is, agents often leave it unread, or read it and still choose another paper (Figure~\ref{fig:exp-dynamics-reason}). In turn, accuracy drops 19 to 37pp, and even the best agent returns a wrong paper for about a third of queries.

On \CI, agents struggle to \textit{reassess} their way to the right paper, with Claude Code (Opus 4.8) and Codex CLI (GPT-5.6-Sol) failing in different ways. Claude Code often leaves a wrong answer but keeps picking new wrong papers, still losing 15pp from the corresponding single-turn queries, while Codex picks a wrong paper far less often in the first place, yet still loses 6. As a result, the two, which score alike on single queries, end 11pp apart (Figure~\ref{fig:exp-dynamics-cum}).

On \NM, agents struggle to \textit{abstain} when no paper qualifies. They open the paper closest to the request within ten tool calls in over 90\% of queries, yet still present a paper as a match in 47\% to 82\% of queries (Figure~\ref{fig:exp-dynamics-no-match}). A researcher may thus mistake an open direction for one already taken.

In summary, our contributions are as follows:
\begin{itemize}[leftmargin=12pt,noitemsep,topsep=2pt]
    \item We introduce \textsc{CompLit} to evaluate scientific literature search on researchers' implicit, cumulative, and unmet needs across eight domains.
    \item We test thirteen baselines, from corpus retrievers to agents, reporting trends by domain and publication year.
    \item We also analyze search dynamics to pinpoint where each agent fails, not only in retrieving a paper but also in verifying it.
\end{itemize}


\section{Related Work}

\begin{table*}[t]
\centering
\footnotesize
\setlength{\tabcolsep}{5pt}
\begin{adjustbox}{width=\linewidth}
\begin{tabular}{@{}l ll ccc >{\scriptsize}c cr@{}}
\toprule
& \multicolumn{2}{c}{Constraint} & \multicolumn{3}{c}{Task} & & & \\
\cmidrule(lr){2-3} \cmidrule(lr){4-6}
\textbf{Benchmark} & Type & Level & Implicit & Cumulative & Unmet & \multicolumn{1}{c}{Domains} & Scope & \textbf{$|Q|$} \\
\midrule
LitSearch \citeyearpar{ajith2024litsearch}         & \ding{172}                               & E\,P & \xmark & \xmark & \xmark & CS & Abs  & 597 \\
PaSa \citeyearpar{he2025pasa}                      & \ding{172}                               & P    & \xmark & \xmark & \xmark & CS & Abs  & 1,050 \\
DORIS-MAE \citeyearpar{wang2023scientific}         & \ding{172}                               & E\,P & \xmark & \xmark & \xmark & CS & Abs  & 100 \\
PaperFindingBench \citeyearpar{bragg2026astabench} & \ding{172}\ding{174}                     & E\,P & \xmark & \xmark & \xmark & CS & Full & 267 \\
LitQA2-Search \citeyearpar{bragg2026astabench}     & \ding{172}                               & P    & \xmark & \xmark & \xmark & BIO & Full & 75 \\
PaSaMaster-Bench \citeyearpar{du2026towards}       & \ding{172}\ding{174}                     & E\,P & \xmark & \xmark & \xmark & CS, EE, PHY, MATH, BIO, ECO & Full & 244 \\
SAGE \citeyearpar{hu2026sage}                      & \ding{172}\ding{173}\ding{174}\ding{175} & E    & \xmark & \xmark & \xmark & CS, PHY, BIO, ECO & Full & 1,200 \\
AutoResearchBench-Deep \citeyearpar{xiong2026autoresearchbench} & \ding{172}\ding{174}\ding{175} & P  & \xmark & \xmark & \cmark & CS & Full & 600 \\
AutoResearchBench-Wide \citeyearpar{xiong2026autoresearchbench} & \ding{172}\ding{174}           & E\,P & \xmark & \xmark & \xmark & CS & Full & 400 \\
\midrule
\texorpdfstring{\CompLitRegular{310}}{CompLit} (Ours) & \ding{172}\ding{173}\ding{174}\ding{175} & \textbf{E\,P\,I} & \cmark & \cmark & \cmark & \textbf{CS, EE, PHY, MATH, BIO, ECO, FIN, STAT} & Full & 870\,+\,220 \\
\bottomrule
\end{tabular}
\end{adjustbox}
\caption{
\textbf{Comparison of \textsc{CompLit} with scientific literature search benchmarks.}
Constraint \textit{Type} is \ding{172} content, \ding{173} figural, \ding{174} metadata, or \ding{175} comparative, a
relation to another paper; \textit{Level} is how the target paper states a constraint, \textbf{E}xplicit,
\textbf{P}araphrased or \textbf{I}nferred, and \textsc{CompLit} alone spans all three.
\textit{Implicit}, \textit{Cumulative} and \textit{Unmet} are three diagnostic tasks.
\textit{Domains} follow arXiv's top-level taxonomy, and \textit{Scope} is the part of a paper needed to answer: an abstract or a full paper. $|Q|$ counts evaluation queries; \textsc{CompLit} has 870 single-turn queries and 220
conversations.
}
\label{tab:benchmark-comparison}
\end{table*}

\paragraph{Literature Search Benchmarks.}
Early benchmarks established natural-language retrieval as a setting for scientific literature search. LitSearch \citep{ajith2024litsearch} and PaSa \citep{he2025pasa} generate queries automatically from citation contexts, each adding a smaller researcher-written set, and PaperFindingBench and LitQA2-Search \citep{bragg2026astabench} draw queries from real user logs and from biology questions. Each of these benchmarks poses a request as a single need rather than a composition of constraints.

\paragraph{Multi-Constraint Queries.}
A complementary line structures complex needs into multiple aspects or conditions. DORIS-MAE \citep{wang2023scientific} decomposes scientific queries into hierarchical aspects, and PaSaMaster-Bench \citep{du2026towards} pairs each expert-written intent with a checklist of criteria across six domains. Beyond scholarly search, BrowseComp \citep{wei2025browsecomp} combines hard-to-find factual web clues, while MultiConIR \citep{lu2025multiconir} scales queries from one to ten conditions. Recent deep-research benchmarks have brought this structure to the scientific literature. SAGE \citep{hu2026sage} composes queries from paper artifacts and citation overlaps, while AutoResearchBench \citep{xiong2026autoresearchbench} derives Deep and Wide queries from target papers. 

By contrast, we start from the researcher's side, where a request requires an agent to \textit{infer} evidence a paper leaves implicit, to \textit{reassess} candidates as the requirements accumulate, and to \textit{abstain} when no paper meets the need.

\paragraph{Implicit, Cumulative, and Unmet.}
Each of our diagnostic settings has a counterpart outside scholarly search. Retrieval and reading comprehension benchmarks make relevance depend on facts a document states only implicitly \citep{taghavi2025impliret, tchuindjo2026obliq}. Conversational benchmarks reveal the need over turns and show that models anchor on early assumptions \citep{ali2026recor,laban2026llms}. Unanswerable questions test whether models abstain \citep{rajpurkar2018know,kirichenko2026abstentionbench}. Scholarly search demands all three as a researcher explores what they have not: properties are rarely stated outright, requirements grow with each paper read, and their conjunction may match no paper. Yet no scholarly benchmark in Table~\ref{tab:benchmark-comparison} tests them. While \citet{xiong2026autoresearchbench} covers abstention, only 60 of its queries do so, all in computer science. \textsc{CompLit} tests each ability in a dedicated setting, and its analysis of search dynamics shows not only whether an agent fails but where in the search it does (\S\ref{sec:trajectories}).

\section{The CompLit Benchmark}
\label{sec:benchmark}

We propose \textsc{CompLit}, a benchmark for compositional scientific literature search.  We define the task (\S\ref{sec:task}), derive the constraints of each query (\S\ref{sec:constraints}), build a \Std and three diagnostic settings (\S\ref{sec:settings}), and provide a corpus (\S\ref{sec:corpus}).

\subsection{Task Definition}
\label{sec:task}

A query $q$ expresses a research intent and a set of constraints $\mathcal{C}=\{c_1,\dots,c_k\}$ that a paper the researcher has not yet found must satisfy jointly. A system searches a corpus $\mathcal{D}$ of papers through a fixed index or the open web. 

A paper $p\in\mathcal{D}$ satisfies $c_i$, written $p \models c_i$, if its full paper, with figures, bibliographic record, or comparison with another paper supports $c_i$, and $p \models \mathcal{C}$ if it satisfies every constraint. The gold answer is the target paper $p^*$ with $p^* \models \mathcal{C}$, or $\varnothing$ if no paper in $\mathcal{D}$ satisfies $\mathcal{C}$. A system returns a single paper or abstains; an answer is correct if it is $p^*$, or an abstention when the gold is $\varnothing$, and we report accuracy on each set.

\vspace{-1mm}
\subsection{Query Generation Pipeline}
\label{sec:constraints}

We use \texttt{Qwen3.5-122B-A10B} \citep{qwen35} as the LLM and the VLM in every step of constraint derivation and query generation, with prompts in Appendix~\ref{app:prompt_templates}.

\vspace{-1mm}
\paragraph{Constraint sources.}
We start with survey comparison tables that record each surveyed paper's attributes. The survey authors consider distinguishing in their field, so their rows share a scope yet differ on those attributes.
Each constraint $c_i$ is a proposition about the target paper with a source $s_i\in\{\text{content}, \text{figural}, \text{metadata},$ $\text{comparative}\}$.

From the table, \emph{content} constraints are the cell values of the paper's row, restated as propositions, and \emph{comparative} constraints contrast the paper with another paper in the same table (e.g., broader imaging modalities). 
From the paper itself, \emph{figural} constraints are claims a vision-language model (VLM) reads from its figures, either quantitative (an ordering, a peak value) or qualitative (an architecture pattern, a trend), and \emph{metadata} constraints are the year, categories, and authors in its arXiv record.

These sources test full-paper understanding, figure interpretation, and cross-paper comparison.

\vspace{-1mm}
\paragraph{Constraint verification.}
To ensure factuality and high quality, we verify every constraint exhaustively against the target paper.
When $s_i\in\{\text{content},\text{comparative}\}$ we do not know which part of the paper supports the constraint, so a \textit{grounder} LLM reads the paper and records the supporting passage, and a \textit{verifier} scores whether that passage appears in the paper and supports the constraint. When $s_i\in\{\text{figural},\text{metadata}\}$, the evidence is already known: a VLM \textit{verifier} scores each figural claim against its figure and caption, and metadata constraints are checked against the arXiv record.

If the verifier scores a constraint below an empirical threshold of 0.9, we reground or regenerate it and score it again for up to 10 rounds; any constraint that never passes is discarded. Every constraint that passes is then rechecked by the authors, so each clears both automatic and manual reviews.

\vspace{-1mm}
\paragraph{Evidence levels.}
We label every content, figural, and comparative constraint with an evidence level $\ell_i\in\{\text{explicit},\text{paraphrased},$ $\text{inferred}\}$, which records how a system must ground it in the paper. We judge it against the paper for a content constraint, and against the figure for a figural one. 

A constraint is \textit{explicit} when the system can read the fact as it stands, and \textit{paraphrased} when the system must match meaning in different words, or when the claim re-describes what the figure directly depicts. It is \textit{inferred} when the system must derive the fact, because the text only implies it (e.g., training in-house), or because it follows only from comparing several visual elements (e.g., a ranking or a trend). 

The \textit{inferred} level also covers every comparative constraint, since establishing one requires reading a second paper. These labels are the basis of the \RI setting (\S\ref{sec:reasoning-intensive}).

\vspace{-2mm}
\paragraph{Query writing.}
We select constraints to fit a research intent, so that a query carries only the properties a researcher would need. An intent is generated from one of 16 search purposes \citep{haddad2026understanding}, such as methodological guidance or comparative analysis. For each paper, we select up to three intents it can serve and verify each with the same LLM. The LLM then chooses the constraints that serve each intent and writes a request around it. We repeat generation and verification to ensure quality, and record the constraints it uses as $\mathcal{C}$.

\vspace{-2mm}
\paragraph{Human verification.}
The authors then check every query three ways: each constraint's recorded evidence must be in the paper and support the constraint; the assigned intent must fit the paper and claim nothing the paper does not support; and the query must identify the target.

\vspace{-2mm}
\paragraph{Uniqueness check.}
We then check automatically that every gold paper is the only known match. For each query, we retrieve the top-50 papers with both BM25 \citep{robertson2009probabilistic} and \texttt{Qwen3-Embedding-8B} \citep{zhang2025qwen3embedding}, and the LLM judges whether any of the 100 papers satisfies all constraints. We remove a query if any paper other than the target does, or if any paper at all does for an \NM query. This removed 5.5\%, 4.3\%, and 1.0\% of queries in \Std, \RI, and \NM, respectively.

\subsection{Diagnostic Settings}
\label{sec:settings}

 The settings follow how a literature search unfolds. Alongside \Std, each setting evaluates one challenge.

\vspace{-1mm}
\paragraph{\Std.} This setting, close to existing benchmarks, serves as a reference for the diagnostic settings. Queries state constraints at all evidence levels, given at once, and every query has a target. This set supports analysis across domains, publication years, and numbers of constraints. It contains 350 queries over 251 target papers.

\vspace{-1mm}
\paragraph{\RI.}
\label{sec:reasoning-intensive} This setting tests whether a system can \textit{infer} that a paper meets requirements it states only implicitly. Every constraint must be inferred: $\ell_i=\text{inferred}$ for every $c_i\in\mathcal{C}$, so queries carry no metadata constraints. No constraint can then be resolved by lexical or semantic matching. Rather, each must be reasoned out from what the paper reports or from how it compares to a named paper. It contains 220 queries over 179 papers.

\vspace{-2mm}
\paragraph{\CI.}
\label{sec:cumulative-intent} This setting tests whether a system \textit{reassesses} its earlier answers as requirements accumulate over turns. The need is revealed over multiple turns: following \cite{laban2026llms}, we rebuild each \RI query into a conversation of 3 to 5 turns. An offline scheduler (gpt-5.6-sol \citep{openai2026gpt56}) partitions $\mathcal{C}$ into disjoint subsets $\mathcal{C}^{(1)}, \dots, \mathcal{C}^{(T)}$, one per turn with one to three constraints each, and writes the user utterance for each turn in advance. 

At evaluation time, a gpt-5.4-mini \citep{openai2026gpt54} user simulator at temperature 0 delivers each utterance, so every system faces the same conversation with the same constraints at each turn \cite{laban2026llms}. The system may answer at every turn and is scored on its final answer, once all of $\mathcal{C}$ has been revealed, with per-turn answers kept for analysis. This set contains 220 conversations.

\vspace{-2mm}
\paragraph{\NM.}
\label{sec:no-match}
This setting tests whether a system \textit{abstains} when no paper meets every requirement.
We construct queries as in \Std and retain those answered correctly by all three reference systems, chosen for their low cost: Harness-1 \citep{jiang2026harness}, DeepSeek-V4-Pro \citep{deepseekai2026v4}, and GLM-5.2 \citep{glm5team2026glm5vibecodingagentic}.
We then add a false constraint $\tilde{c}$ by perturbing a verified constraint of the target that is not in $\mathcal{C}$, such as substituting another paper's survey table value or reversing a figure's trend, so that no paper in $\mathcal{D}$ satisfies $\mathcal{C}\cup\{\tilde{c}\}$ and the gold is $\varnothing$.
Since every target is thus easy to find, this setting separates abstention from retrieval. It contains 300 queries.

\subsection{Corpus} 
\label{sec:corpus}

For index-based search, we provide a corpus of 2.95M full-text arXiv papers from the S2ORC v2 snapshot \citep{lo2020s2orc}, with survey papers removed so that corpus-based systems cannot rely on them to find the target, while keeping every target paper. Open-web systems are not restricted, but rarely open the source survey (Appendix~\ref{app:corpus}).

\section{Experiments}

\vspace{-1mm}
We compare thirteen baselines on the four settings (\S\ref{sec:main-results}), then analyze accuracy by domain and publication year on \Std (\S\ref{sec:standard-breakdown}) and search dynamics on \RI, \CI, and \NM (\S\ref{sec:trajectories}).

\subsection{Experimental Setup}
\label{sec:experimental-setup}

\paragraph{Evaluation sets and corpus.}
We use the four settings in \S\ref{sec:settings}: \Std, \RI, \CI, and \NM, with 350, 220, 220, and 300 instances. Corpus-based baselines search the 2.95M-paper collection described in \S\ref{sec:corpus}. 

\vspace{-2mm}
\paragraph{Baselines.} We evaluate thirteen baselines in three groups by how they acquire candidate papers. \textbf{(i) Corpus retrieval}: BM25 \citep{robertson2009probabilistic}, SPECTER2 \citep{singh2023scirepeval}, ReasonIR \citep{shao2025reasonir}, and Qwen3-Embed-8B \citep{zhang2025qwen3embedding} alone and with Qwen3-Reranker-8B. \textbf{(ii) Corpus search agents}: DCI-Agent-Lite \citep{li2026beyond}, which explores the raw corpus via shell, and Harness-1 \citep{jiang2026harness}, which searches it with a state-externalizing harness. \textbf{(iii) Web search agents}: DR Tulu \citep{shao2026drtulu}, PaSa \citep{he2025pasa}, DeepSeek-V4-Pro \citep{deepseekai2026v4}, Kimi-K3 \citep{kimiteam2026k3}, Claude Code \citep{anthropic2025claudecode} with Opus 4.8, and Codex CLI \citep{openai2026gpt56} with GPT-5.6-Sol, which search the open web; we shorten the last four to DeepSeek, Kimi, Claude, and Codex, and refer to them as the frontier agents.

Corpus retrieval always returns a paper and cannot abstain, so we exclude it from \NM. \CI requires holding a conversation across turns, which only DeepSeek, Kimi, Claude, and Codex support, so we run it on these four.

\vspace{-2mm}
\paragraph{Tools in web agents.}
DR Tulu and PaSa run with their released pipelines. Kimi and DeepSeek run in \texttt{inspect\_evals} with three tools: \texttt{arxiv\_search} with arXiv API, \texttt{web\_search} through Serper, and \texttt{read\_paper}, which returns a paper's full text with figures, given a message limit of 30. Claude and Codex run in their own web search harnesses, plus \texttt{arxiv\_search} and \texttt{read\_paper} as MCP tools.  

\vspace{-2mm}
\paragraph{Metrics.}
Each baseline either returns a paper or abstains, since the task is to find one paper that satisfies all constraints jointly (\S\ref{sec:task}). Corpus retrieval answers with its top-ranked paper (i.e., Recall@1), so it is scored comparably to agents. Agents end each run with a JSON object holding a title and a rationale, and an empty title indicates abstention. 

\begin{figure*}[t!]
    \centering
    \includegraphics[width=1\linewidth]{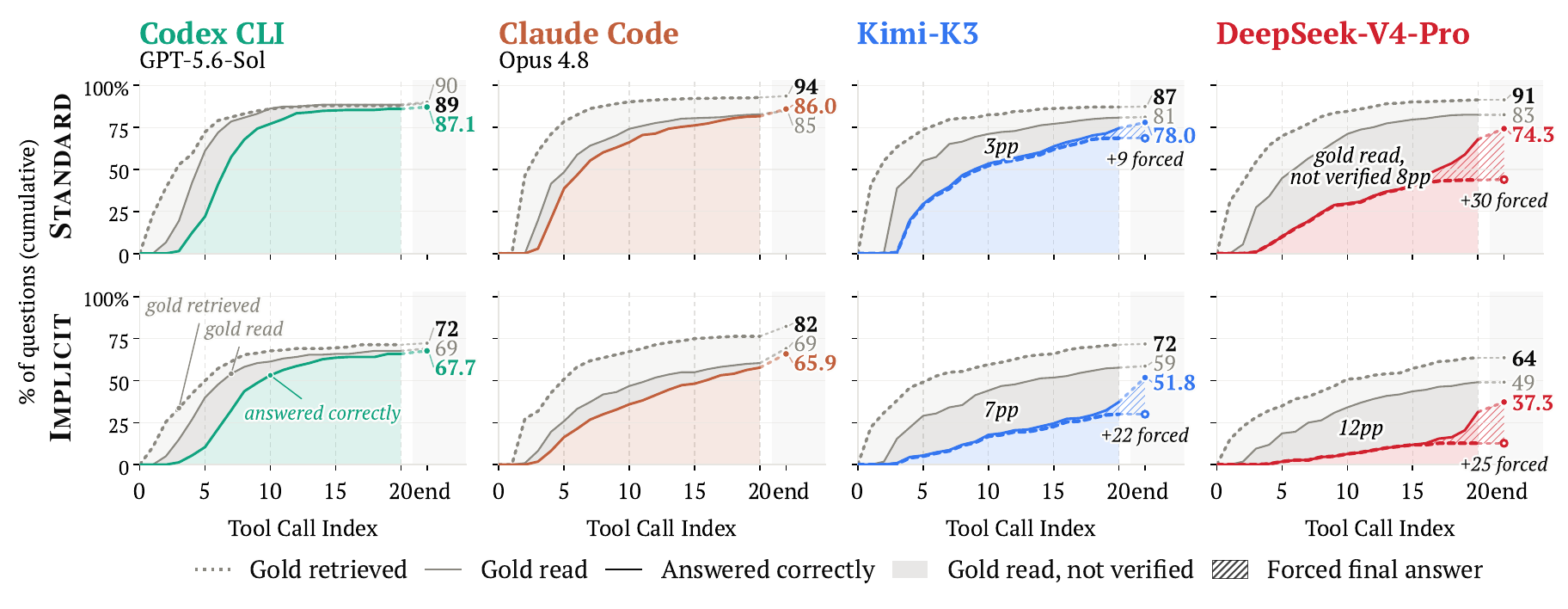} 
    \vspace{-8mm}
    \caption{
    \textbf{Search dynamics on \Std\ and \RI.} By tool call, the share of queries in which the gold paper has been retrieved (dotted), read (solid gray), and answered correctly (colored). The \emph{end} column shows final values. Hatching marks answers that Kimi and DeepSeek give only when forced at the end of the budget.
    }
    \vspace{-3mm}
    \label{fig:exp-dynamics-reason}
\end{figure*}

We report accuracy, where a prediction is correct if it matches the gold paper's title, or if the baseline abstains on an \NM\ query. On \CI\ we score the answer at the final turn. Kimi and DeepSeek runs that do not commit within the message limit receive a forced final answer, counted toward accuracy and marked separately in Figure~\ref{fig:exp-dynamics-reason}.

\begin{table}[t]
\centering
\footnotesize
\setlength{\tabcolsep}{2.5pt}
\renewcommand{\arraystretch}{1.1}
\begin{tabularx}{\columnwidth}{@{}Xcccc@{}}
\toprule
Baseline & \textsc{Std.} & \textsc{Imp.} & \textsc{Cum.} & \textsc{Unm.} \\
\midrule
\rowcolor{retrievalrow}
\multicolumn{5}{c}{\textit{Corpus Retrieval}} \\
SPECTER2           & 4.6  & 0.1  & -- & -- \\
BM25               & 28.9 & 3.2  & -- & -- \\
ReasonIR           & 15.4 & 5.5  & -- & -- \\
Qwen3-Embed-8B     & 49.1 & 18.6 & -- & -- \\
\quad + Qwen3-Rerank-8B & \bestscore{66.0} & \bestscore{25.9} & -- & -- \\
\addlinespace[2pt]
\rowcolor{corpusagentrow}
\multicolumn{5}{c}{\textit{Corpus Search Agents}} \\
DCI-Agent-Lite$^\dagger$ & 54.3 & 24.1 & -- & 5.9 \\
Harness-1          & \bestscore{69.7} & \bestscore{42.3} & -- & \bestscore{9.0} \\
\addlinespace[2pt]
\rowcolor{webagentrow}
\multicolumn{5}{c}{\textit{Web Search Agents}} \\
DR Tulu            & 23.7 & 7.3  & -- & 14.3 \\
PaSa               & 42.6 & 15.9 & -- & 8.7 \\
DeepSeek-V4-Pro    & 74.3 & 37.3 & 35.5 & 17.3 \\
Kimi-K3            & 78.0 & 51.8 & 37.3 & 18.7 \\
Claude Code (Opus 4.8)  & 86.0 & 65.9 & 50.5 & 26.3 \\
Codex CLI (GPT-5.6-Sol) & \bestscore{87.1} & \bestscore{67.7} & \bestscore{61.8} & \bestscore{52.7} \\
\bottomrule
\end{tabularx}
\vspace{-2mm}
\caption{
  \textbf{Main results on \textsc{CompLit}.}
  Accuracy (\%) on \Std, \RI, \CI, and \NM (abbreviated as shown).
  On \CI, the final-turn answer is scored; on \NM, accuracy is the share of queries on which the baseline abstains.
  -- marks settings a baseline does not support (\S\ref{sec:experimental-setup}).
  $^\dagger$With GPT-5.4 nano.
  The best score in each group is in \textbf{bold}.
}
\vspace{-3mm}
\label{tab:main-results}
\end{table}

\vspace{-2mm}
\subsection{Main Results}
\label{sec:main-results}

\paragraph{On \Std, the four frontier agents score high, while the rest vary widely within each group and no group is consistently ahead.} The frontier agents score 74.3\% to 87.1\%, with Codex and Claude one point apart. The rest spread widely: corpus retrieval ranges from 4.6\% to 66.0\%. Interestingly, the reranked retriever, at 66.0\%, sits between the two corpus search agents and is even 23pp above PaSa. This set shows which baselines are effective within each group, but not how they can fail when constraints are inferred, cumulative, or unsatisfiable.

\vspace{-1mm}
\paragraph{On \RI, all baselines degrade, and even the strongest by about 20pp.} Codex and Claude degrade by 19 and 20pp, Kimi by 26, and DeepSeek by 37, showing that the weaker agents degrade more. The reranked retriever degrades the most, by 40pp, whereas the corpus search agents, which browse the same index, degrade by 27--30. This suggests that agents' reasoning over papers helps establish properties that must be inferred.

\vspace{-1mm}
\paragraph{On \CI, Codex and Claude, which score alike on a single query, end 11pp apart.}
All four agents degrade relative to \RI, by 2 to 15pp. In particular, the gap between Codex and Claude widens from 2pp to 11 as constraints are imposed over turns. This suggests that a single-query evaluation can obscure the gap between agents in handling cumulative requirements.

\paragraph{On \NM, most baselines return a paper, and even the strongest fails to abstain on almost half of the queries.} Codex abstains correctly on 52.7\% of queries and Claude on 26.3\%, while every other baseline stays below 20\%. Among them, Harness-1, DR Tulu, and PaSa, which are trained on answerable queries, abstain on 9.0\%, 14.3\%, and 8.7\%. This suggests that verifying every constraint and abstaining if even one fails must be trained explicitly.

\colorlet{accH}{clBlue!70}   
\colorlet{accG}{clBlue!60}   
\colorlet{accF}{clBlue!50}   
\colorlet{accE}{clBlue!30}   
\colorlet{accD}{clBlue!10}   
\colorlet{accC}{clBlue!8}    
\colorlet{accB}{clBlue!6}    
\colorlet{accA}{clBlue!4}    
\colorlet{accZ}{clBlue!2}    
\newcommand{\acc}[1]{%
  \ifdim #1pt<60pt \cellcolor{accZ}\else
  \ifdim #1pt<65pt \cellcolor{accA}\else
  \ifdim #1pt<70pt \cellcolor{accB}\else
  \ifdim #1pt<75pt \cellcolor{accC}\else
  \ifdim #1pt<80pt \cellcolor{accD}\else
  \ifdim #1pt<85pt \cellcolor{accE}\else
  \ifdim #1pt<90pt \cellcolor{accF}\else
  \ifdim #1pt<95pt \cellcolor{accG}\else
  \cellcolor{accH}\fi\fi\fi\fi\fi\fi\fi\fi
  #1}
\begin{table}[t]
    \centering
    \small
    \setlength{\tabcolsep}{5pt}
    \begin{tabular}{@{}lcccc@{}}
    \toprule
    & Codex & Claude & Kimi & DeepSeek \\
    \midrule
    \multicolumn{5}{@{}l}{\textit{Domain}} \\
    Physics & \acc{87.3} & \acc{79.4} & \acc{68.3} & \acc{60.3} \\
    EESS & \acc{88.3} & \acc{83.3} & \acc{80.0} & \acc{76.7} \\
    Statistics & \acc{84.9} & \acc{90.6} & \acc{67.9} & \acc{66.0} \\
    Computer science & \acc{80.4} & \acc{84.3} & \acc{84.3} & \acc{70.6} \\
    Biology & \acc{88.4} & \acc{90.7} & \acc{81.4} & \acc{83.7} \\
    Mathematics & \acc{87.5} & \acc{82.5} & \acc{77.5} & \acc{80.0} \\
    Finance \& economics & \acc{95.0} & \acc{95.0} & \acc{92.5} & \acc{92.5} \\
    \midrule
    \multicolumn{5}{@{}l}{\textit{Publication year}} \\
    $\leq$2015 & \acc{79.5} & \acc{78.2} & \acc{65.4} & \acc{71.8} \\
    2016--18 & \acc{87.3} & \acc{84.1} & \acc{73.0} & \acc{63.5} \\
    2019--21 & \acc{92.6} & \acc{96.3} & \acc{87.7} & \acc{81.5} \\
    2022--23 & \acc{93.8} & \acc{89.2} & \acc{81.5} & \acc{80.0} \\
    $\geq$2024 & \acc{82.5} & \acc{81.0} & \acc{82.5} & \acc{73.0} \\
    \bottomrule
    \end{tabular}
    
    \vspace{-2mm}
    \caption{
        \textbf{Accuracy by domain and publication year of the target.} Results are for the four frontier agents on \Std. Darker cells mark higher accuracy.
    }
    \vspace{-5mm}
    \label{tab:standard-breakdown}
\end{table}

\subsection{Accuracy by Domain and Year}
\label{sec:standard-breakdown}

Covering all eight arXiv domains and papers through 2026, \textsc{CompLit} can show researchers in each field which agent serves them well and where it may fall short. We break down the \Std\ accuracy of the four frontier agents by domain and publication year (Table~\ref{tab:standard-breakdown}).

\vspace{-1mm}
\paragraph{No agent leads across all domains, and each agent struggles with different domains.} Codex leads in physics, EESS, and mathematics, and Claude in statistics and biology. Codex scores 85\% to 88\% in most domains but 80.4\% in computer science, below Claude and Kimi. Claude scores above 90\% in statistics and biology but 79.4\% in physics. Kimi and DeepSeek are weakest in physics and statistics, and DeepSeek is also the lowest of the four in EESS and computer science. 

\vspace{-1mm}
\paragraph{All four are more accurate on papers in 2019--23 than on older or more recent ones.} Relative to their best period, the agents score 14 to 22pp lower on papers from 2018 or earlier: Codex, Claude, and Kimi are weakest on papers from 2015 or earlier, and DeepSeek on those from 2016 to 2018. On papers from 2024 onward, Codex, Claude, and DeepSeek score 8 to 15pp below their best, and Kimi 5pp below. This shows that agents are less reliable on older and very recent papers, which may be harder to retrieve or verify.

\begin{figure*}[t!] 
\centering
\includegraphics[width=1\linewidth]{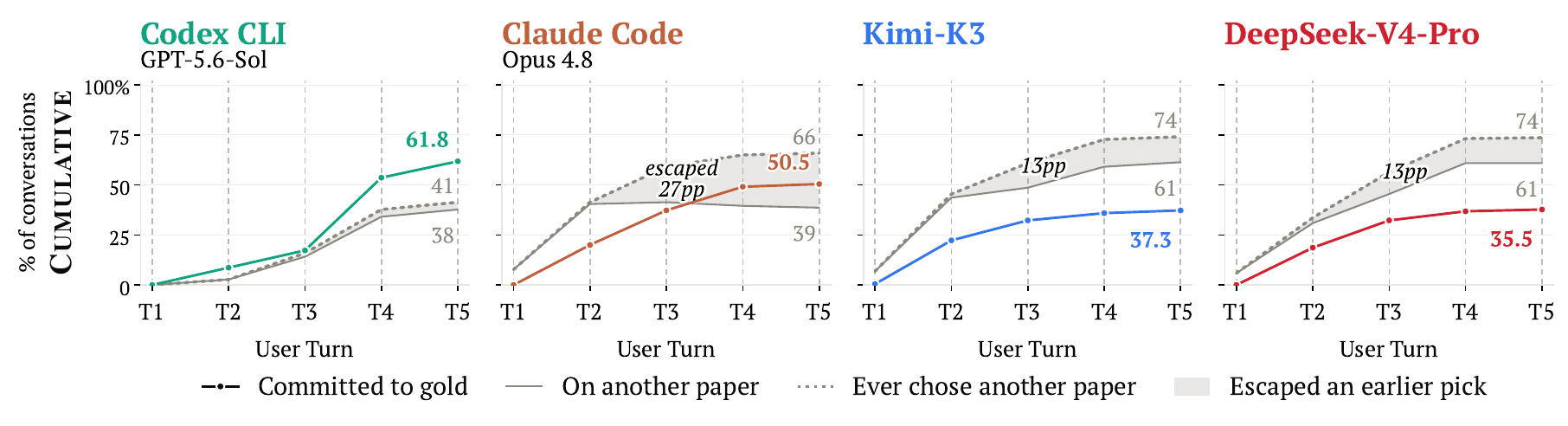} 
    \vspace{-8mm}
    \caption{
    \textbf{Revision dynamics on \CI.} By user turn, the share of conversations whose current answer is the gold paper (colored) or another paper (solid gray), and the share that has chosen another paper at some turn (dotted). The shaded band shows conversations that chose a wrong paper at an earlier turn but later escaped it.
    }
    \vspace{-2mm}
    \label{fig:exp-dynamics-cum}
\end{figure*}

\begin{figure*}[t!]
    \centering
    \includegraphics[width=1\linewidth]{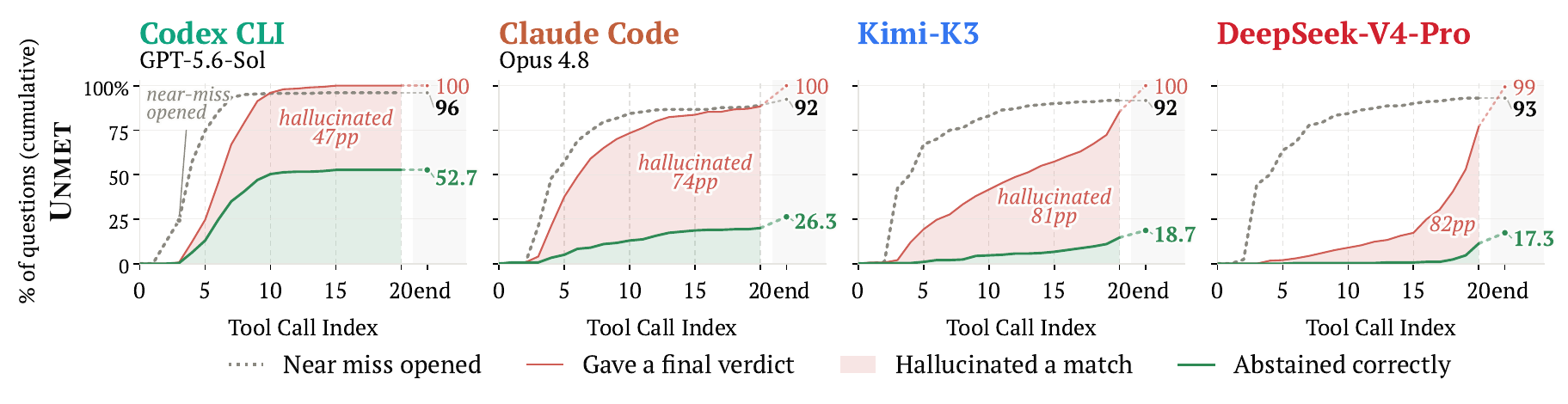} 
    \vspace{-8mm}
    \caption{
    \textbf{Abstention dynamics on \NM.} By tool call, the share of queries in which the agent has opened the near-miss paper (dotted), has given a final verdict (red), and has abstained correctly (green). The shaded area shows verdicts that name a paper although no paper satisfies all constraints.
    }
    \vspace{-4mm}
    \label{fig:exp-dynamics-no-match}
\end{figure*}

\subsection{Analysis of Search Dynamics}
\label{sec:trajectories}

\subsubsection{
\Std vs. \RI
}

Figure~\ref{fig:exp-dynamics-reason} tracks, over tool calls, the share of queries in which the gold paper has appeared in search results (\emph{retrieved}),\footnote{Codex does not expose its native web search results, so we count only \texttt{arxiv\_search} and \texttt{read\_paper} calls.} has been opened (\emph{read}), and has been given as the answer (\emph{answered correctly}), on \Std\ and \RI.

\paragraph{On \RI, agents fail more often at every step, across retrieving, reading, and answering with the gold.}
In Figure~\ref{fig:exp-dynamics-reason}, all three curves shift down and to the right from \Std\ to \RI: at the same number of tool calls, the gold paper is retrieved, read, or answered in fewer queries. The final share of queries with the gold paper \textit{retrieved} falls by 12 to 27pp, the share with it \textit{read} by 16 to 34pp, and accuracy by 19 to 37pp. 

The gaps between the steps also widen, both from \textit{retrieved} to \textit{read} and from \textit{read} to \textit{answered correctly}. This suggests that when every constraint must be inferred, agents find it harder both to decide which paper to read and to verify every constraint in a paper they have read.

\paragraph{The weaker the agent, the wider the gaps between retrieving, reading, and answering with the gold paper.} On \RI, the gap between retrieving and reading the gold paper is 3pp for Codex and 13 to 15pp for the other three agents, and the gap between reading and answering grows from 1 point for Codex to 3, 7, and 12pp for Claude, Kimi, and DeepSeek. Also, Kimi and DeepSeek answer many of the papers they read only when forced, recovering 22 and 25pp of their accuracy. This suggests that agents have room to improve not only in retrieving the gold paper but also in reading and answering with it once retrieved.

\subsubsection{Revision on \CI}

Figure~\ref{fig:exp-dynamics-cum} tracks, by user turn, the share of conversations whose current answer is the gold paper (\emph{committed to gold}) or another paper (\emph{on another paper}), the share that has chosen another paper at some turn (\emph{ever chose another}), and their difference (\emph{escaped}).

\paragraph{Across turns, agents differ in how they revise their answers, and wrong answers accumulate for the weaker agents.} Two questions separate the agents: whether wrong answers rise faster than correct ones, and if so, whether the agent escapes them. For Codex, correct answers rise faster, leaving few wrong ones to escape. For the other three, wrong answers lead from the second turn. Claude escapes many of them, so the share of conversations currently on a non-gold paper (solid gray) remains near 40\%, even though the share that has ever chosen one (dotted) continues to rise. Kimi and DeepSeek escape a few, and their current wrong share rises to 61\%. These differences emerge only as requirements arrive over turns, so a single-turn evaluation cannot show whether an agent revises its earlier answers.

\subsubsection{Abstention on \NM} 

Each \NM\ query has a \emph{near-miss}, the original target that satisfies every constraint except the injected one. Figure~\ref{fig:exp-dynamics-no-match} tracks, by tool call, whether the agent has opened the near-miss (\emph{near-miss opened}) and given its final answer (\emph{gave a final verdict}). That answer either names a paper, the near-miss or any other, as satisfying every constraint (\emph{hallucinated a match}), or abstains (\emph{abstained correctly}).

\paragraph{All agents open the near-miss early, but most go on to accept it as a match.}
All four agents open the near-miss in over 90\% of queries within the first ten tool calls. Yet only Codex abstains in about half of the queries, while the other three end with a paper, nearly always the near-miss, in 74\% to 82\% of queries. Since \NM\ keeps only queries whose target the reference baselines already find (\S\ref{sec:no-match}), the results suggest that agents fail at verification: they find the near-miss but do not abstain.

\paragraph{Agents open the near-miss at the same point, but decide at different ones.}
Codex commits within a few calls of opening it, whether it abstains or not, while Claude and Kimi decide gradually and DeepSeek mostly in its last few calls. The later an agent decides, the less often it abstains, from 52.7\% for Codex to 17.3\% for DeepSeek. Stronger agents thus seem to better gauge their confidence in the near-miss \citep{yoon2025reasoning}.
\section{Conclusion}
We introduced \textsc{CompLit}, a benchmark that evaluates agents in the way researchers search: outward from what they have read, with needs that are implicit, cumulative, or unmet. Even the strongest agents lose about 20 points when every constraint is implicit and fall further, each in a different way, as requirements accumulate, while most agents accept a near-miss instead of abstaining. Beyond final accuracy, \textsc{CompLit} traces each agent's search step by step, revealing where these failures arise: agents often reach the right paper, or the closest one, yet misjudge it. An agent that does well here can be trusted to find the paper that meets a need, or to say that none does.

Looking ahead, we envision two directions. One is a harness for literature search, guided by where each agent fails and which papers it finds hard. The other is a world model of a field, distilled from surveys, so that an agent knows how to search a field and check a condition before the search begins. We hope \textsc{CompLit} serves as a foundation for future work on scientific literature search.

\section*{Limitations}
\textsc{CompLit} draws its constraints from comparison tables in arXiv surveys from English arXiv papers and relies on the distinctions survey authors chose to compare. Constraints and queries are generated and verified automatically, but we check every constraint and every query by hand to remove remaining errors. Finally, open-web agents search a live web, so their results may shift slightly if the experiments are rerun.


\bibliography{custom}

\clearpage
\appendix
\section{Benchmark Construction}
\label{app:construction}

\subsection{Benchmark Statistics}
\label{app:stats}

\paragraph{Overview.}
\textsc{CompLit} draws on 193 survey papers that contain at least one comparison table over the papers they review. Table~\ref{tab:app-overview} summarizes each set. The four sets cover 443 distinct target papers. \Std and \RI share no target paper, and \CI reuses the 220 queries of \RI. \NM shares 142 target papers with \Std and 41 with \RI, but no query.

\begin{table}[h]
\centering
\footnotesize
\setlength{\tabcolsep}{4pt}
\begin{tabular}{@{}lrrrr@{}}
\toprule
 & \textsc{Std.} & \textsc{Imp.} & \textsc{Cum.} & \textsc{Unm.} \\
\midrule
Queries or conversations & 350 & 220 & 220 & 300 \\
Target papers & 251 & 179 & 179 & 196 \\
Surveys & 115 & 117 & 117 & 102 \\
Constraints per query & 6.9 & 5.2 & 5.2 & 7.8 \\
Turns per conversation & -- & -- & 3.9 & -- \\
\bottomrule
\end{tabular}
\caption{\textbf{Overview of the four sets.} Settings are abbreviated as in Table~\ref{tab:main-results}. \NM counts include the injected false constraint.}
\label{tab:app-overview}
\end{table}

\begin{figure*}[t]
\centering
\includegraphics[width=\linewidth]{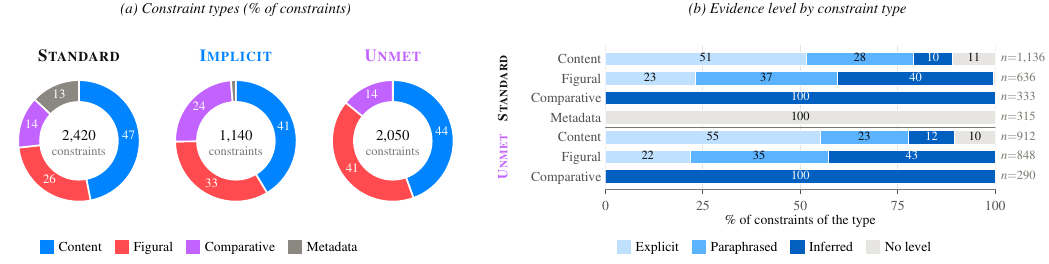}
\vspace{-5mm}
\caption{\textbf{Constraints in \textsc{CompLit}.} (a) Share of each constraint type, over all constraints of a set. \CI has the constraints of \RI, and \NM excludes the injected constraint. (b) Evidence level of each constraint type in \Std and \NM. Comparative constraints are inferred by definition, and metadata constraints carry no level. \textit{No level} also marks content constraints whose field was renamed during query writing and no longer matches a labeled attribute.}
\label{fig:app-composition}
\end{figure*}

\paragraph{Constraint types and evidence levels.}
Figure~\ref{fig:app-composition} shows how constraints divide by type and level. Content constraints are the largest type in every set, and figural constraints make up a quarter of \Std and two fifths of \NM. In \Std, about half of the content constraints are explicit, while figural constraints spread across all three levels. With comparative constraints counted as inferred, 28.9\% of \Std constraints are inferred. \RI draws only from inferred content constraints, inferred figural claims whose fact the paper's text does not state, and comparative constraints, so its levels are fixed by construction.

\begin{figure*}[t]
\centering
\includegraphics[width=\linewidth]{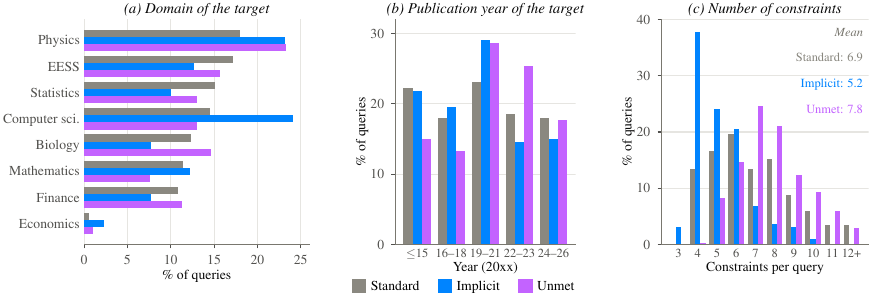}
\vspace{-5mm}
\caption{\textbf{Targets and queries in \textsc{CompLit}.} (a) arXiv group of the target paper's primary category, where physics pools \texttt{astro-ph}, \texttt{cond-mat}, \texttt{hep}, \texttt{nucl}, \texttt{quant-ph}, and \texttt{physics}. (b) Publication year of the target. (c) Number of constraints per query, where \NM includes the injected constraint. \CI has the distributions of \RI.}
\label{fig:app-distribution}
\end{figure*}

\paragraph{Domains, years, and query length.}
Figure~\ref{fig:app-distribution} shows the spread of targets and queries. Every set covers all eight arXiv groups. Physics is the largest group in \Std and \NM, computer science in \RI, and economics the smallest in all three. Target papers range from 1998 to 2026, with a median year of 2020 in \Std and 2019 in \RI. \Std queries carry 4 to 17 constraints, while \RI queries carry 3 to 10, most often four.

\begin{figure}[t]
\centering
\includegraphics[width=\linewidth]{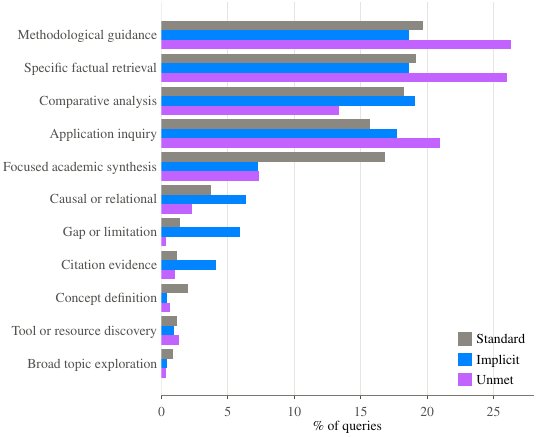}
\vspace{-6mm}
\caption{\textbf{Intent of each query}, from 11 search purposes of \citet{haddad2026understanding}. \CI has the intents of \RI at the query level.}
\label{fig:app-intents}
\end{figure}

\paragraph{Intents.}
Intents follow the taxonomy of \citet{haddad2026understanding}. We keep the 11 of its 16 purposes that ask for papers and drop the 5 generative ones, such as drafting and ideation. Figure~\ref{fig:app-intents} gives their distribution: five purposes account for over 80\% of queries in every set, and \RI draws more on causal, gap, and citation purposes than \Std.

\subsection{Human Verification}
\label{app:human}

The authors reviewed every paper in a web interface that shows the paper's full text and figures next to each generated item. Review proceeds in stages, and a paper enters the benchmark only after it passes the query stage.

\begin{enumerate}[leftmargin=*,itemsep=1pt,topsep=2pt]
\item \textbf{Constraints.} For each content, comparative, and figural constraint, the reviewer checks that the recorded evidence appears in the paper and supports the constraint, and corrects its value or level. Verifier-dropped table cells are shown too, so a wrong drop can be restored.
\item \textbf{Intents.} The reviewer checks that each intent fits the paper and claims nothing the paper does not support.
\item \textbf{Queries.} The reviewer checks that the query identifies the target and that its constraint set matches the text, and edits either one.
\end{enumerate}

Table~\ref{tab:app-human} counts the changes over the 360 papers that passed the query stage of the main review track. Reviewers removed between a fifth and a quarter of the content, comparative, and figural constraints and edited the values of a quarter of the remaining content constraints. They also restored 120 table cells the verifier had dropped and edited 81 of them. Few intents and queries were removed outright. Most query edits changed the constraint set to match the text. \RI was reviewed on a separate track with the same three checks, where 26 of 400 queries were removed and 86 had their constraint set changed.

\begin{table}[h]
\centering
\footnotesize
\setlength{\tabcolsep}{3pt}
\begin{tabular}{@{}lrrr@{}}
\toprule
Item & Before & Removed & Edited \\
\midrule
Content constraints & 1,582 & 304 (19.2\%) & 317 \\
Comparative constraints & 849 & 215 (25.3\%) & 18 \\
Figural claims & 4,654 & 851 (18.3\%) & 1,378 \\
Intents & 894 & 2 (0.2\%) & 333 \\
Queries & 2,480 & 1 (0.0\%) & 137 \\
\bottomrule
\end{tabular}
\caption{\textbf{Changes made in human verification}, over the 360 papers of the main review track. \textit{Edited} counts items whose value or text changed; for queries, it counts changed constraint sets. The figural count is net of 228 claims that reviewers split or rewrote.}
\label{tab:app-human}
\end{table}

\subsection{\CI and \NM Construction}
\label{app:ci-un}

\begin{figure}[t]
\centering
\includegraphics[width=\linewidth]{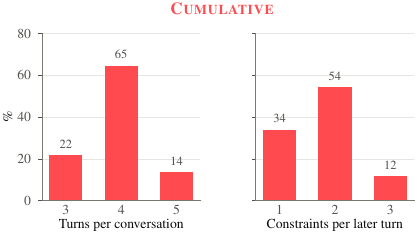}
\vspace{-6mm}
\caption{\textbf{\CI conversations.} Turns per conversation and constraints per turn after the opening, as a share of the 220 conversations and of their 642 later turns.}
\label{fig:app-ci-nm}
\end{figure}

\paragraph{Scheduler rules.}
The \CI scheduler turns each \RI query into a conversation under the following rules, and code enforces every rule except the last.
\begin{itemize}[leftmargin=*,itemsep=1pt,topsep=2pt]
\item The opening turn states the researcher's broad goal and carries no constraint. Its motivation stays one level more general than every constraint, and the audit caps its score when it shares two or more content words with them.
\item Each later turn carries one to three constraints that form one thought, and every constraint appears exactly once.
\item A conversation has at least three turns, and the last turn asks the agent to commit to one paper or to say that none qualifies.
\item Each turn names a search purpose from the intent catalog, and the purpose may change across turns.
\item The scheduler holds back the constraint that most directly identifies the target for a later turn, so the opening alone does not identify it. This rule is given only as an instruction.
\end{itemize}
An audit by \texttt{gpt-5.6-sol} rescores each turn's motivation and asks for a rewrite below 0.8. Figure~\ref{fig:app-ci-nm} summarizes the resulting conversations: most have four turns, and most later turns reveal two constraints.

\paragraph{User simulator.}
At evaluation time, \texttt{gpt-5.4-mini} at temperature 0 delivers the prepared utterance of each turn. When the agent's last message asks a question or proposes a paper, the simulator adds one sentence that answers from what the user has already said, or that states the project has no requirement on that point. It never adds a constraint, so every agent faces the same constraints at each turn.

\paragraph{\NM construction.}
We start from answerable queries that the reference systems answer correctly, so that a failure on \NM reflects abstention rather than retrieval. We use only the variant without metadata. For each query, \texttt{gpt-5.6-sol} writes one false clause of one of two kinds.
\begin{itemize}[leftmargin=*,itemsep=1pt,topsep=2pt]
\item \textbf{A table value.} The clause asserts a false value on a survey table column that the query does not use. The true value must appear in the paper's body but not in its title or abstract.
\item \textbf{A figure fact.} The clause reverses a verified figural claim, such as an ordering or a trend, whose fact the paper's text also states.
\end{itemize}
Every original sentence of the query is kept, and the clause adds 7 to 30 words. A judge then reads the target paper and labels the clause as \textit{refuted} when the paper contradicts it, \textit{silent} when the paper does not settle it, or \textit{satisfied}. We drop satisfied clauses.

\subsection{Examples}
\label{app:examples}

We show one instance per setting, with every constraint, its type, its evidence level, and the evidence that supports it in the target paper (Figures~\ref{fig:ex-standard}--\ref{fig:ex-unmet}). Metadata constraints carry no level (\S\ref{sec:constraints}), and a comparative constraint is always \clevel{inferred}. The released \RI records do not store a level for most constraints, so the levels we show for the \RI example come from our own reading of the paper.

\begin{figure*}[p]
\begin{examplebox}[colStandard]{Example of a \Std query (stat)}
\textbf{Query.} I'm building a drug-target binding affinity predictor and need a clean dual-CNN design that processes compound and protein inputs through separate convolutional streams before merging them. The method I'm implementing requires exactly this two-branch architecture with concatenation prior to the fully connected layers, so I can reproduce the pipeline without wrestling with graph-based or recurrent alternatives. I need the work that uses the same type of ligand input as `WideDTA: prediction of drug-target binding affinity' but a different protein representation from `DeepAffinity: Interpretable Deep Learning of Compound-Protein Affinity through Unified Recurrent and Convolutional Neural Networks'. It comes from the late 2010s on arXiv.

\smallskip
\textbf{Intent.} Methodological guidance. \quad
\textbf{Gold.} \textit{DeepDTA: Deep Drug-Target Binding Affinity Prediction} (arXiv:1801.10193, 2018). \quad
\textbf{Survey.} \textit{Drug-Target Interaction/Affinity Prediction: Deep Learning Models and Advances Review} (arXiv:2502.15346).

\smallskip
\renewcommand{\arraystretch}{1.1}
\begin{tabularx}{\linewidth}{@{}l l L{0.28\linewidth} X@{}}
\toprule
Type & Level & Constraint & Evidence in the target paper \\
\midrule
\ctype{content} & \clevel{explicit} & a CNN encodes the drug & \evid{We proposed a CNN-based prediction model that comprises two separate CNN blocks, each of which aims to learn representations from SMILES strings and protein sequences.} (Proposed Model) \\
\ctype{content} & \clevel{explicit} & a CNN encodes the protein & the same sentence \\
\ctype{figural} & \clevel{paraphrased} & two streams with separate CNN blocks for compound SMILES and protein sequences & Figure~2, \evid{DeepDTA model with two CNN blocks to learn from compound SMILES and protein sequences.} \\
\ctype{figural} & \clevel{paraphrased} & a concatenation merges the two streams before the fully connected layers & Figure~2 draws the concatenation feeding the fully connected layers, and the text states \evid{The final features of the max-pooling layers were concatenated and fed into three FC layers} (Proposed Model). \\
\ctype{comparative} & \clevel{inferred} & the same ligand representation as WideDTA (arXiv:1902.04166) & The survey table records SMILES strings as the ligand input of both papers, with WideDTA also adding substructure words. DeepDTA predates WideDTA and never mentions it. \\
\ctype{comparative} & \clevel{inferred} & a different protein representation than DeepAffinity (arXiv:1806.07537) & The survey table records protein sequences for DeepDTA and structural property strings for DeepAffinity. DeepDTA never mentions DeepAffinity. \\
\ctype{metadata} & -- & year $=$ 2018; venue $=$ arXiv & arXiv record \\
\bottomrule
\end{tabularx}
\end{examplebox}
\vspace{-3mm}
\caption{\textbf{A \Std query.} Its eight constraints cover all four types and all three evidence levels. The query names WideDTA and DeepAffinity but not their input representations, so the two comparative constraints require the survey table or the two papers.}
\label{fig:ex-standard}
\end{figure*}

\begin{figure*}[p]
\begin{examplebox}[colImplicit]{Example of an \RI query (math)}
\textbf{Query.} I'm trying to identify parametric spatiotemporal dynamical models where the coefficients exhibit step changes, but I'm struggling with the sharp transitions in the loss landscape when tuning regularization. I need a procedure that handles these coefficient variations without requiring Bayesian uncertainty quantification, and I want to avoid methods that get bogged down in handling missing data. I also need an approach that is illustrated with less real data than `DeepMoD: Deep learning for Model Discovery in noisy data', so I can focus on the core identification pipeline without that overhead.

\smallskip
\textbf{Intent.} Methodological guidance. \quad
\textbf{Gold.} \textit{Data-driven identification of parametric partial differential equations} (arXiv:1806.00732, 2018). \quad
\textbf{Survey.} \textit{A Review of Data-Driven Discovery for Dynamic Systems} (arXiv:2210.10663).

\smallskip
\renewcommand{\arraystretch}{1.1}
\begin{tabularx}{\linewidth}{@{}l l L{0.25\linewidth} X@{}}
\toprule
Type & Level & Constraint (survey cell) & Evidence in the target paper \\
\midrule
\ctype{content} & \clevel{inferred} & no Bayesian uncertainty quantification (UQ $=$ no) & The paper never mentions Bayes, posteriors, or uncertainty. It selects one model per method with a point criterion: \evid{we evaluate the models using the AIC-inspired loss function} (\S2.3). \\
\ctype{content} & \clevel{inferred} & does not handle missing data (missing data $=$ no) & The paper never mentions missing data. Every example is simulated, and \evid{For the noise-free dataset we use the discrete Fourier transform for computing derivatives} (\S3.1). \\
\ctype{figural} & \clevel{inferred} & the loss stays low over the regularization sweep, then jumps & Figure~2 plots the loss for 50 thresholds and 50 penalties, and in both panels it stays near its minimum and then jumps. The text says only \evid{For sufficiently high values, the model is too sparse and is no longer predictive} (\S2.3). \\
\ctype{figural} & \clevel{paraphrased} & the coefficients show a step change & The Figure~5 caption states it: \evid{coefficients for Navier-Stokes equations exhibiting jump in Reynolds number from 100 to 75 at t = 10.} \\
\ctype{comparative} & \clevel{inferred} & less real data than arXiv:1904.09406 & The gold is \evid{demonstrated on four canonical models}, all simulated, while the other paper is applied to \evid{noisy experimental time-series data from a gel electrophoresis experiment}. The gold predates it and never mentions it. \\
\bottomrule
\end{tabularx}
\end{examplebox}
\vspace{-3mm}
\caption{\textbf{An \RI query.} Four of the five constraints must be inferred: the paper never states that it is non-Bayesian or ignores missing data, the jump in the loss is visible only in Figure~2, and the comparative constraint names a property without its value. The step change is stated in the text and in the Figure~5 caption. The loss-curve constraint also separates the gold from Online WSINDy (arXiv:2203.03979), a later method that satisfies the other four constraints but never plots loss against its threshold. The agents' answers to this query appear in Table~\ref{tab:qual-implicit-math}.}
\label{fig:ex-implicit}
\end{figure*}

\begin{figure*}[p]
\begin{examplebox}[colCumulative]{Example of a \CI conversation (stat)}
\textbf{Gold.} \textit{Model-based Differentially Private Data Synthesis and Statistical Inference in Multiply Synthetic Differentially Private Data} (arXiv:1606.08052, 2016). \quad
\textbf{Survey.} \textit{Comparative Study of Differentially Private Data Synthesis Methods} (arXiv:1602.01063, revised through 2019).

\smallskip
\renewcommand{\arraystretch}{1.15}
\begin{tabularx}{\linewidth}{@{}c L{0.2\linewidth} X@{}}
\toprule
Turn & Revealed & User utterance \\
\midrule
1 & none (opening) & I'm developing a critique of approaches that protect sensitive information while still allowing useful data release, and I want a concrete limitation to anchor the argument. Can you help me find one paper that provides that case? \\
2 & \ctype{content} $\times$2: privacy cost only on sanitizing the posterior; releasing records is free post-processing & The first thing I need is a clean accounting boundary, since I don't want to blame any step that consumes no additional privacy budget. The privacy cost should fall only on sanitizing the posterior, with generation of the released records treated as free post-processing. \\
3 & \ctype{content} $\times$2: one route needs an intractable normalizing constant; the other an analytic sensitivity bound & With that boundary in place, the bottleneck has to persist across both natural routes for carrying out the protected inference, so it cannot be dismissed as an implementation quirk. One route must face an intractable normalizing constant, while the other must require an analytic sensitivity bound. Please commit to one paper now, or say plainly if nothing meets everything I've asked for. \\
\bottomrule
\end{tabularx}
\end{examplebox}
\vspace{-3mm}
\caption{\textbf{A \CI conversation.} The opening states the researcher's goal and field but no constraint. The next two turns reveal two constraints each, and the agent is scored on its answer after the last turn. The gold states that drawing the released records \evid{does not incur any additional privacy cost as it can be regarded as post-processing}, notes that the posterior's \evid{normalizing constant \dots\ might not have a close-form expression}, and sanitizes sufficient statistics with the Laplace mechanism, which requires their global sensitivity.}
\label{fig:ex-cumulative}
\end{figure*}

\begin{figure*}[p]
\begin{examplebox}[colUnmet]{Example of an \NM query (stat)}
\textbf{Query.} I need a Python tool that implements a structured four-step causal inference workflow, taking domain knowledge and data to construct causal graphs and estimate effects. \textcolor{colUnmet!80!black}{\textbf{The workflow should identify the target estimand before modeling the causal mechanisms.}} It should support standard identification strategies like backdoor, frontdoor, and instrumental variable adjustments, with built-in mechanisms to refute estimates and validate results before deployment.

\smallskip
\textbf{Gold.} $\varnothing$ (abstain). \quad
\textbf{Near-miss.} \textit{DoWhy: An End-to-End Library for Causal Inference} (arXiv:2011.04216, 2020). \quad
\textbf{Intent.} Tool and resource discovery. \quad
\textbf{Perturbation.} A figure fact, reordered.

\smallskip
\renewcommand{\arraystretch}{1.1}
\begin{tabularx}{\linewidth}{@{}l l L{0.30\linewidth} X@{}}
\toprule
Type & Level & Constraint & Status in the near-miss \\
\midrule
\ctype{content} & \clevel{explicit} & implemented in Python & satisfied \\
\ctype{content} & \clevel{paraphrased} & supports backdoor, frontdoor, and IV identification & satisfied \\
\ctype{figural} & \clevel{paraphrased} & a four-step analysis pipeline & satisfied (Figure~1) \\
\ctype{figural} & \clevel{explicit} & the four steps are Model causal mechanisms, Identify the target estimand, Estimate causal effect, and Refute estimate & satisfied (Figure~1); not stated in the query, and the injected clause reverses its first two steps \\
\ctype{figural} & -- & \textcolor{colUnmet!80!black}{\textbf{identify the estimand before modeling causal mechanisms}} (injected) & \textcolor{badRed}{violated}: \evid{DoWhy's API is organized around the four key steps that are required for any causal analysis: Model, Identify, Estimate, and Refute.} \\
\bottomrule
\end{tabularx}
\end{examplebox}
\vspace{-3mm}
\caption{\textbf{An \NM query.} The injected constraint (bold) reverses the step order that DoWhy's Figure~1 depicts. The original target thus becomes a near-miss that satisfies every other constraint, and the correct answer is to abstain.}
\label{fig:ex-unmet}
\end{figure*}

\clearpage

\section{Lexical Retrieval Across Benchmarks}
\label{app:bm25-benchmarks}

We compare how hard \textsc{CompLit} questions are to match lexically against existing literature-search benchmarks. We run one BM25 retriever over one corpus for six benchmarks and for \Std and \RI.

\paragraph{Setup.}
The corpus holds the titles and abstracts of all 3,182,867 arXiv papers in the arXiv metadata snapshot of September 28, 2026. Unlike the BM25 baseline of Appendix~\ref{app:baselines}, it indexes no full bodies, so that the gold papers of every benchmark are in the corpus. BM25 uses \texttt{tantivy} ($k_1{=}1.2$, $b{=}0.75$) and takes each question verbatim. Several benchmarks list many gold papers per question, which caps Recall@1 below one, so we report H@$k$, the share of questions with at least one gold paper in the top $k$.

\paragraph{Gold papers.}
We map every gold paper to an arXiv id and skip questions left with none. PaSa~\citep{he2025pasa} lists arXiv ids for RealScholarQuery and the AutoScholarQuery test split. For LitSearch~\citep{ajith2024litsearch}, 520 of 639 gold papers have an arXiv id in Semantic Scholar, which leaves 498 of 597 questions. For PaperFindingBench~\citep{bragg2026astabench}, we use the May 2025 test split; 392 of 803 listed papers have an arXiv id, and most semantic questions list no gold paper because they are scored by a judge, which leaves 101 of 267 questions. For AutoResearchBench~\citep{xiong2026autoresearchbench}, we use the Wide subset. For DORIS-MAE~\citep{wang2023scientific}, we use the BIRCO release, match 5,505 of its 5,544 candidate abstracts to arXiv by text, and count a paper as gold if its relevance score is at least 1.0. DORIS-MAE is thus scored against all of arXiv rather than its 100-paper candidate pools. We exclude LitQA2, since only one of its 199 questions cites an arXiv paper.

\paragraph{Results.}
\textbf{Among benchmarks with one or two gold papers per question, \RI is the hardest to match at the top ranks, while \Std is the easiest (Table~\ref{tab:app-bm25-benchmarks}).} BM25 ranks a gold paper first for 1.8\% of \RI questions, against 7.1--18.5\% on PaSa, PaperFindingBench, and LitSearch. By the top 100 the gap closes, so the gold paper is often retrieved but rarely ranked first: the constraints it satisfies are inferred rather than stated in the question. \Std reaches 25.7\% at H@1, the highest of all benchmarks, consistent with its constraints being stated in the question. PaperFindingBench includes 27 metadata questions on authors and years, which a title-and-abstract index cannot answer.

\begin{table}[t!]
    \centering
    \small
    \setlength{\tabcolsep}{3pt}
    \begin{tabular}{@{}lcccc@{}}
    \toprule
    Benchmark & H@1 & H@5 & H@20 & H@100 \\
    \midrule
    PaSa (RealScholarQuery) & 12.0 & 34.0 & 50.0 & 74.0 \\
    PaSa (AutoScholarQuery) & 7.1 & 15.1 & 25.8 & 41.2 \\
    PaperFindingBench & 14.9 & 26.7 & 36.6 & 44.6 \\
    LitSearch & 18.5 & 30.9 & 42.0 & 56.2 \\
    AutoResearchBench (Wide) & 14.2 & 36.0 & 61.5 & 84.5 \\
    DORIS-MAE & 0.0 & 25.0 & 62.5 & 83.9 \\
    \midrule
    CompLit \Std & 25.7 & 45.4 & 58.0 & 74.6 \\
    CompLit \RI & 1.8 & 13.2 & 27.3 & 48.2 \\
    \bottomrule
    \end{tabular}
    \caption{
        \textbf{Lexical retrieval across literature-search benchmarks.} BM25 over title and abstract of a shared corpus of 3.18M arXiv papers. H@$k$ is the share of questions with at least one gold paper in the top $k$. 
    }
    \label{tab:app-bm25-benchmarks}
\end{table}

\section{Does the Agent Harness Matter?}
\label{app:harness}

Claude Code and Codex CLI run in their own harnesses, while Kimi and DeepSeek run in \texttt{inspect\_evals}. We test whether the harness, rather than the model, explains part of the gap, by running DeepSeek on \RI inside the Claude Code and Codex CLI harnesses.

\paragraph{Setup.}
The model, tools, and prompt are those of the \texttt{inspect\_evals} run (Appendix~\ref{app:baselines}): DeepSeek-V4-Pro, the same \texttt{arxiv\_search}, text-only \texttt{read\_paper}, and Serper \texttt{web\_search} exposed as MCP tools, and the same single user prompt. The built-in web tools of both harnesses are disabled. Each harness keeps its own system prompt, agent loop, and context management. Without a limit, DeepSeek keeps searching and rarely answers, so each harness allows 30 tool calls per query, after which further calls are blocked with an instruction to answer. In \texttt{inspect\_evals}, a query ends after 30 messages (median 19 tool calls), and runs that have not committed receive a forced final answer.

\paragraph{Results.}
\textbf{The two harnesses score the same, and both match \texttt{inspect\_evals} once an answer is forced (Table~\ref{tab:app-harness}).} DeepSeek reaches 41.4\% in Claude Code and 40.9\% in Codex CLI, against 37.3\% in \texttt{inspect\_evals} with the forced final answer. Without it, \texttt{inspect\_evals} reaches 12.7\%, so the gap comes from forcing a commitment, not from the harness. The remaining 4pp coincides with the larger tool budget of the harnesses (30 against a median of 19 calls). In the same harnesses, DeepSeek stays far below Claude Code and Codex CLI with their own models (65.9\% and 67.7\%, Table~\ref{tab:main-results}), so the harness does not explain that gap.

\begin{table}[t]
    \centering
    \small
    \setlength{\tabcolsep}{5pt}
    \begin{tabular}{@{}lccc@{}}
    \toprule
    Harness & Forced answer & Tool calls & Acc. \\
    \midrule
    \texttt{inspect\_evals} & -- & 19 & 12.7 \\
    \texttt{inspect\_evals} & \checkmark & 19 & 37.3 \\
    Claude Code & \checkmark & 31 & 41.4 \\
    Codex CLI & \checkmark & 30 & 40.9 \\
    \bottomrule
    \end{tabular}
    \caption{
        \textbf{DeepSeek-V4-Pro on \RI in three harnesses.} Model, tools, and prompt are fixed. Tool calls is the median per query. In \texttt{inspect\_evals}, the answer is forced after the 30-message limit; in Claude Code and Codex CLI, after 30 tool calls.
    }
    \label{tab:app-harness}
\end{table}

\clearpage
\onecolumn
\subsection{Prompt Templates}
\label{app:prompt_templates}

We list the core prompts for constraint derivation and query writing, verbatim, except that we cut passages not needed to follow them and mark each cut as \texttt{[... omitted ...]}. Placeholders in braces are filled per paper. We omit the verifier and retry prompts that accompany each step: every verifier scores the generator's output against the same inputs, and a retry appends the verifier's issues to the generator prompt. Each step runs \texttt{Qwen3.5-122B-A10B} at temperature 0.3 with the system message \textit{``You are a helpful research assistant. Respond only with valid JSON.''}

\paragraph{Content constraints: the grounder.}
Given the survey's table cells for the target paper and up to 60,000 characters of the paper's body, the grounder gives each cell a readable attribute name, keeps its value, quotes the supporting passage, labels its evidence level, and keeps or drops it. The labels \textit{literal}, \textit{semantic}, and \textit{inferential} correspond to explicit, paraphrased, and inferred.
\promptfile[colImplicit]{Prompt for the Grounder (content constraints and their evidence levels)}{tex/appendix/prompts/core_grounder.txt}

\paragraph{Comparative constraints.}
Given the whole comparison table, the LLM writes two to five facts that each distinguish the target from one named sibling paper on one column.
\promptfile[colImplicit]{Prompt for Comparative Constraints}{tex/appendix/prompts/core_comparative.txt}

\paragraph{Figural constraints.}
The VLM receives one figure with its caption and writes at most three claims about what the figure shows. A few papers were processed with other Qwen3.5 sizes.
\promptfile[colImplicit]{Prompt for Figural Claims (VLM)}{tex/appendix/prompts/core_figural.txt}

\paragraph{Query writing.}
Given an intent and the paper's verified constraints, the LLM selects the constraints that serve the intent, writes the request, and records the constraints it used.
\promptfile[colImplicit]{Prompt for Query Writing}{tex/appendix/prompts/core_query.txt}

\twocolumn

\clearpage
\section{Experimental Details}
\label{app:setup}
\subsection{Baseline Configurations}
\label{app:baselines}

We describe the configuration of every baseline. All runs took place between August 12 and August 24, 2026. Web agents searched the live web in that window.

\paragraph{Corpus retrieval.}
Every retriever returns a top-100 ranking over the 2.95M-paper corpus, and we score its first paper (Recall@1). BM25 indexes titles and full bodies with \texttt{tantivy} ($k_1{=}1.2$, $b{=}0.75$). SPECTER2 embeds the title and the first 2,000 characters of the body. ReasonIR and Qwen3-Embed-8B embed up to 32,512 tokens per paper, and Qwen3-Embed-8B uses the instruction \textit{``Given a web search query, retrieve relevant passages that answer the query''}. Qwen3-Rerank-8B reranks the top 100 of Qwen3-Embed-8B using each paper's title and opening, up to 4,096 tokens.

\paragraph{Corpus search agents.}
DCI-Agent-Lite runs \texttt{gpt-5.4-nano} with high thinking effort and \texttt{read} and \texttt{bash} tools over per-month files of the arXiv metadata snapshot, which hold titles, authors, abstracts, and categories, for at most 100 turns and one hour per query. Harness-1 is served with vLLM and searches the corpus with a hybrid of BM25 and Qwen3-Embed-8B, with grep, document reading, and pruning tools. It runs a single rollout with at most 50 steps and a 32,768-token context, at temperature 1.0.

\paragraph{Open-web agents.}
DR Tulu and PaSa run their released pipelines. DR Tulu-8B uses Google search through Serper, Semantic Scholar snippet search, and a page browser summarized by Qwen3-8B, with at most 10 tool calls. PaSa's crawler generates five search queries and then expands two citation layers, with at most 20 papers per layer. Kimi and DeepSeek run in \texttt{inspect\_ai} as a ReAct agent with no system message, provider-default decoding, and three tools:
\begin{itemize}[leftmargin=*,itemsep=1pt,topsep=2pt]
\item \texttt{arxiv\_search} queries the arXiv API and returns titles, ids, years, authors, and truncated abstracts;
\item \texttt{web\_search} queries Google through Serper and returns titles, URLs, and snippets;
\item \texttt{read\_paper} returns a paper's full text and attaches up to four figures as images. DeepSeek reads no images.
\end{itemize}
Claude Code and Codex CLI use their built-in web search and receive \texttt{arxiv\_search} and \texttt{read\_paper} as MCP tools. Claude may also call \texttt{WebFetch}, and Codex a shell restricted to reading files. They have no turn limit and a wall-clock limit of 900 seconds per query.

\subsection{Corpus and Survey Access}
\label{app:corpus}

Each query is built from a survey's comparison table, so its source survey lists the target with its attributes. We remove these surveys from the corpus for index-based search (\S\ref{sec:corpus}), but open-web systems can still reach them. We therefore count how often the four frontier agents open the source survey of a query, either with \texttt{read\_paper} or by fetching its page.

Across the four settings, the frontier agents open the source survey in 3.8\% of runs (164 of 4,360): 1.8\% for Codex, 4.3\% for Claude, 4.9\% for Kimi, and 4.0\% for DeepSeek. Opening it does not help. Accuracy on queries where the agent opens the source survey is lower than on the rest in 10 of the 16 agent and setting pairs, and each of the other six rests on 12 or fewer such queries. This suggests that agents turn to the survey when a query is already hard, and that it does not lead them to the target.

\section{Additional Results}
\label{app:results}

This section extends the analysis of \S\ref{sec:standard-breakdown} and \S\ref{sec:trajectories} to every baseline with per-query results. These are the four frontier agents, the two corpus search agents, DR Tulu, and PaSa. The corpus retrievers are omitted because only their aggregate scores were kept.

\subsection{Accuracy by Domain and Year}
\label{app:domain-year}

Table~\ref{tab:app-domain-year} extends Table~\ref{tab:standard-breakdown} to eight baselines and to \RI. On \Std, the pattern of the four frontier agents extends to the other baselines. Every baseline except DeepSeek is least accurate on papers from 2015 or earlier, and physics is the hardest domain for six of the eight. On \RI, the domain ranking changes: mathematics becomes the hardest domain for Claude and Codex, both at 48.1\%, while statistics becomes one of their easiest.

\begin{table*}[t]
\centering
\scriptsize
\setlength{\tabcolsep}{3.5pt}
\begin{tabular}{@{}lr cccccccc c r cccccccc@{}}
\toprule
& \multicolumn{9}{c}{\Std} & & \multicolumn{9}{c}{\RI} \\
\cmidrule(lr){2-10} \cmidrule(lr){12-20}
& $n$ & DCI & H-1 & Tulu & PaSa & DS & Kimi & Claude & Codex & & $n$ & DCI & H-1 & Tulu & PaSa & DS & Kimi & Claude & Codex \\
\midrule
\multicolumn{20}{@{}l}{\textit{Domain}} \\
Physics & 63 & \acc{46.0} & \acc{57.1} & \acc{15.9} & \acc{22.2} & \acc{60.3} & \acc{68.3} & \acc{79.4} & \acc{87.3} & & 51 & \acc{15.7} & \acc{33.3} & \acc{3.9} & \acc{7.8} & \acc{29.4} & \acc{49.0} & \acc{72.5} & \acc{66.7} \\
EESS & 60 & \acc{53.3} & \acc{75.0} & \acc{26.7} & \acc{41.7} & \acc{76.7} & \acc{80.0} & \acc{83.3} & \acc{88.3} & & 28 & \acc{32.1} & \acc{39.3} & \acc{7.1} & \acc{10.7} & \acc{28.6} & \acc{39.3} & \acc{50.0} & \acc{64.3} \\
Statistics & 53 & \acc{58.5} & \acc{75.5} & \acc{18.9} & \acc{43.4} & \acc{66.0} & \acc{67.9} & \acc{90.6} & \acc{84.9} & & 22 & \acc{27.3} & \acc{36.4} & \acc{13.6} & \acc{22.7} & \acc{68.2} & \acc{59.1} & \acc{90.9} & \acc{81.8} \\
Computer science & 51 & \acc{54.9} & \acc{64.7} & \acc{37.3} & \acc{41.2} & \acc{70.6} & \acc{84.3} & \acc{84.3} & \acc{80.4} & & 53 & \acc{22.6} & \acc{39.6} & \acc{3.8} & \acc{22.6} & \acc{35.8} & \acc{47.2} & \acc{62.3} & \acc{64.2} \\
Biology & 43 & \acc{58.1} & \acc{76.7} & \acc{23.3} & \acc{58.1} & \acc{83.7} & \acc{81.4} & \acc{90.7} & \acc{88.4} & & 17 & \acc{23.5} & \acc{64.7} & \acc{5.9} & \acc{11.8} & \acc{47.1} & \acc{70.6} & \acc{64.7} & \acc{82.4} \\
Mathematics & 40 & \acc{55.0} & \acc{62.5} & \acc{17.5} & \acc{42.5} & \acc{80.0} & \acc{77.5} & \acc{82.5} & \acc{87.5} & & 27 & \acc{29.6} & \acc{59.3} & \acc{11.1} & \acc{14.8} & \acc{25.9} & \acc{48.1} & \acc{48.1} & \acc{48.1} \\
Finance \& economics & 40 & \acc{57.5} & \acc{80.0} & \acc{27.5} & \acc{60.0} & \acc{92.5} & \acc{92.5} & \acc{95.0} & \acc{95.0} & & 22 & \acc{27.3} & \acc{40.9} & \acc{13.6} & \acc{22.7} & \acc{45.5} & \acc{68.2} & \acc{77.3} & \acc{81.8} \\
\midrule
\multicolumn{20}{@{}l}{\textit{Publication year}} \\
$\leq$2015 & 78 & \acc{44.9} & \acc{57.7} & \acc{11.5} & \acc{28.2} & \acc{71.8} & \acc{65.4} & \acc{78.2} & \acc{79.5} & & 48 & \acc{22.9} & \acc{39.6} & \acc{10.4} & \acc{14.6} & \acc{31.2} & \acc{47.9} & \acc{79.2} & \acc{72.9} \\
2016--18 & 63 & \acc{52.4} & \acc{77.8} & \acc{17.5} & \acc{34.9} & \acc{63.5} & \acc{73.0} & \acc{84.1} & \acc{87.3} & & 43 & \acc{23.3} & \acc{48.8} & \acc{7.0} & \acc{18.6} & \acc{41.9} & \acc{60.5} & \acc{62.8} & \acc{74.4} \\
2019--21 & 81 & \acc{65.4} & \acc{79.0} & \acc{27.2} & \acc{54.3} & \acc{81.5} & \acc{87.7} & \acc{96.3} & \acc{92.6} & & 64 & \acc{25.0} & \acc{46.9} & \acc{6.2} & \acc{10.9} & \acc{37.5} & \acc{51.6} & \acc{56.2} & \acc{67.2} \\
2022--23 & 65 & \acc{50.8} & \acc{64.6} & \acc{30.8} & \acc{50.8} & \acc{80.0} & \acc{81.5} & \acc{89.2} & \acc{93.8} & & 32 & \acc{31.2} & \acc{40.6} & \acc{6.2} & \acc{18.8} & \acc{43.8} & \acc{50.0} & \acc{84.4} & \acc{75.0} \\
2024--26 & 63 & \acc{57.1} & \acc{69.8} & \acc{33.3} & \acc{44.4} & \acc{73.0} & \acc{82.5} & \acc{81.0} & \acc{82.5} & & 33 & \acc{18.2} & \acc{30.3} & \acc{6.1} & \acc{21.2} & \acc{33.3} & \acc{48.5} & \acc{51.5} & \acc{45.5} \\
\bottomrule
\end{tabular}
\caption{\textbf{Accuracy (\%) by domain and publication year of the target, for every baseline with per-query results.} H-1 is Harness-1, Tulu is DR Tulu, and DS is DeepSeek. Darker cells mark higher accuracy, with the bins of Table~\ref{tab:standard-breakdown}.}
\label{tab:app-domain-year}
\end{table*}

\subsection{Accuracy by Constraint Type and Evidence Level}
\label{app:type-level}

Table~\ref{tab:app-type-level} splits \Std accuracy by the constraints a query contains. For seven of the eight baselines, queries with a comparative constraint are the hardest, and for seven of the eight, queries with a figural constraint are the easiest. Whether a query contains any inferred constraint matters little, but the share of inferred constraints does. When at least half of a query's labeled constraints are inferred, the frontier agents lose 5.7 to 10.1 points against queries with no inferred constraint, and PaSa loses 13.0. This is the effect \RI isolates at full strength, where every constraint is inferred.

\begin{table}[h]
\centering
\scriptsize
\setlength{\tabcolsep}{2.4pt}
\begin{tabular}{@{}lrcccccccc@{}}
\toprule
& $n$ & DCI & H-1 & Tulu & PaSa & DS & Kimi & Cl. & Cx. \\
\midrule
\multicolumn{10}{@{}l}{\textit{Contains a constraint of type}} \\
Content & 349 & 54.4 & 69.6 & 23.8 & 42.7 & 74.5 & 77.9 & 86.0 & 87.1 \\
Figural & 246 & 59.3 & 78.0 & 25.2 & 50.0 & 79.3 & 85.4 & 91.9 & 91.9 \\
Metadata & 178 & 56.2 & 71.9 & 30.3 & 44.4 & 75.8 & 80.3 & 87.6 & 86.0 \\
Comparative & 168 & 48.8 & 66.1 & 22.0 & 42.9 & 69.0 & 71.4 & 83.9 & 83.3 \\
\midrule
\multicolumn{10}{@{}l}{\textit{Share of inferred constraints}} \\
0 & 136 & 51.5 & 68.4 & 17.6 & 44.1 & 72.8 & 75.0 & 86.0 & 90.4 \\
$(0, 0.5)$ & 153 & 58.2 & 73.2 & 29.4 & 45.8 & 79.1 & 84.3 & 88.2 & 86.9 \\
$[0.5, 1)$ & 61 & 50.8 & 63.9 & 23.0 & 31.1 & 65.6 & 68.9 & 80.3 & 80.3 \\
\midrule
\multicolumn{10}{@{}l}{\textit{Number of constraints}} \\
4 & 47 & 53.2 & 57.4 & 14.9 & 27.7 & 70.2 & 68.1 & 70.2 & 80.9 \\
5--6 & 127 & 56.7 & 65.4 & 18.9 & 40.2 & 74.0 & 78.0 & 86.6 & 88.2 \\
7--8 & 100 & 48.0 & 77.0 & 29.0 & 47.0 & 73.0 & 78.0 & 87.0 & 90.0 \\
9+ & 76 & 59.2 & 75.0 & 30.3 & 50.0 & 78.9 & 84.2 & 93.4 & 85.5 \\
\bottomrule
\end{tabular}
\caption{\textbf{\Std accuracy (\%) by the constraints a query contains.} The share of inferred constraints is over content and figural constraints with a level label. Cl.\ is Claude and Cx.\ is Codex.}
\label{tab:app-type-level}
\end{table}

Queries with nine or more constraints are easier than queries with four for every baseline in the table, although accuracy does not rise monotonically in between. Claude rises from 70.2\% on queries with four constraints to 93.4\% on queries with nine or more.

\subsection{\CI by Turn}
\label{app:ci-turns}

Table~\ref{tab:app-ci-turns} gives the numbers behind Figure~\ref{fig:exp-dynamics-cum}. Codex commits late: at the third turn it answers the gold in only 17.3\% of conversations, against 37.3\% for Claude, but it ends 11 points higher. Every agent is more accurate on longer conversations.

\begin{table}[h]
\centering
\scriptsize
\setlength{\tabcolsep}{3.2pt}
\begin{tabular}{@{}lccccccc@{}}
\toprule
& \multicolumn{5}{c}{On gold after turn} & Ever & Recovered \\
\cmidrule(lr){2-6}
Agent & 1 & 2 & 3 & 4 & 5 & another & \\
\midrule
Codex & 0.0 & 8.6 & 17.3 & 53.6 & 61.8 & 41.4 & 3.6 \\
Claude & 0.0 & 20.0 & 37.3 & 49.1 & 50.5 & 65.9 & 25.5 \\
Kimi & 0.5 & 22.3 & 32.3 & 35.9 & 37.3 & 74.1 & 12.3 \\
DeepSeek & 0.0 & 18.6 & 32.3 & 36.8 & 37.7 & 73.6 & 11.8 \\
\midrule
& \multicolumn{3}{c}{Final, by turns} & & & & \\
\cmidrule(lr){2-4}
& 3 & 4 & 5 & & & & \\
\midrule
Codex & 56.2 & 62.0 & 70.0 & & & & \\
Claude & 37.5 & 52.8 & 60.0 & & & & \\
Kimi & 31.2 & 38.0 & 43.3 & & & & \\
DeepSeek & 29.2 & 34.5 & 50.0 & & & & \\
\bottomrule
\end{tabular}
\caption{\textbf{\CI by turn (\%).} \textit{On gold after turn} carries a finished conversation's last answer forward. \textit{Ever another} is the share that chose a non-gold paper at some turn, and \textit{Recovered} the share that did so but ended on the gold. \textit{Recovered} differs from the escaped band of Figure~\ref{fig:exp-dynamics-cum}, which also counts conversations that leave a wrong paper without ending on the gold. The lower block gives final accuracy for conversations of 3, 4, and 5 turns ($n{=}48$, 142, 30).}
\label{tab:app-ci-turns}
\end{table}

\subsection{What Agents Answer on \NM}
\label{app:nm-answers}

Table~\ref{tab:app-nm-answers} splits each baseline's \NM answers into abstentions, the near-miss, and another paper. When they do not abstain, all baselines except DR Tulu name the near-miss more often than all other papers combined. Codex abstains more often when the paper contradicts the injected clause than when it leaves the clause unsettled, 55.2\% against 40.4\%. For the other three agents the two rates differ by less than 8 points.

\begin{table}[h]
\centering
\scriptsize
\setlength{\tabcolsep}{3pt}
\begin{tabular}{@{}lccc cc@{}}
\toprule
& \multicolumn{3}{c}{Answer} & \multicolumn{2}{c}{Abstained, by clause} \\
\cmidrule(lr){2-4} \cmidrule(lr){5-6}
Baseline & Abst. & Near & Other & Ref. & Sil. \\
\midrule
Harness-1 & 9.0 & 67.7 & 23.3 & -- & -- \\
DR Tulu & 14.3 & 29.7 & 56.0 & -- & -- \\
PaSa & 8.7 & 54.3 & 37.0 & -- & -- \\
DeepSeek & 17.3 & 75.3 & 6.7 & 17.7 & 15.4 \\
Kimi & 18.7 & 76.0 & 5.3 & 17.3 & 25.0 \\
Claude & 26.3 & 62.0 & 11.7 & 26.6 & 25.0 \\
Codex & 52.7 & 44.7 & 2.7 & 55.2 & 40.4 \\
\bottomrule
\end{tabular}
\caption{\textbf{\NM answers (\%).} \textit{Abst.}\ is accuracy, \textit{Near} names the near-miss, and \textit{Other} names another paper. On the right, abstention for clauses the paper refutes or leaves silent. DeepSeek's remaining 0.7\% gave no answer. For Harness-1, which returns up to five papers, \textit{Near} and \textit{Other} refer to its first paper.}
\label{tab:app-nm-answers}
\end{table}

\subsection{Cost}
\label{app:cost}

Table~\ref{tab:app-cost} reports the mean cost per query of the frontier agents. \RI and \CI take more tool calls and time than \Std for every agent. Codex is the fastest and makes the fewest tool calls, while Kimi spends the most time in every setting.

\begin{table}[h]
\centering
\footnotesize
\setlength{\tabcolsep}{4pt}
\begin{tabular}{@{}lcccc@{}}
\toprule
Agent & \textsc{Std.} & \textsc{Imp.} & \textsc{Cum.} & \textsc{Unm.} \\
\midrule
Codex & 6.9 / 44 & 9.5 / 72 & 11.4 / 125 & 5.9 / 37 \\
Claude & 8.6 / 135 & 15.5 / 305 & 13.0 / 446 & 9.4 / 152 \\
Kimi & 10.0 / 237 & 16.4 / 371 & 18.1 / 643 & 12.1 / 228 \\
DeepSeek & 13.6 / 143 & 17.7 / 192 & 30.0 / 395 & 16.7 / 199 \\
\bottomrule
\end{tabular}
\caption{\textbf{Mean cost per query}, as tool calls / seconds. Mean dollar cost is \$0.39, \$0.61, \$0.73, and \$0.32 per setting for Codex, computed from its tokens at list price, and \$0.64, \$1.10, \$1.81, and \$0.59 for Claude, as reported by Claude Code.}
\label{tab:app-cost}
\end{table}

\subsection{Qualitative Comparison of Agents}
\label{app:qualitative}
\renewcommand{\topfraction}{0.95}\renewcommand{\textfraction}{0.05}\setcounter{topnumber}{3}

Tables~\ref{tab:qual-standard}--\ref{tab:qual-unmet-math} put the four frontier agents side by side on eight queries from six domains, two for each setting. \textit{Calls} counts tool calls, and \textit{Opened} records whether the agent read the gold paper, or the near-miss on \NM. Excerpts are verbatim from the agents' answers, and we checked every wrong answer against the text of the paper it names. We chose these cases to illustrate how agents differ, among queries where the agents disagree and every wrong answer demonstrably violates a constraint; they are not a random sample, and aggregate accuracy is given in Table~\ref{tab:main-results}.

\paragraph{\Std: checking the one distinctive requirement.}
In Table~\ref{tab:qual-standard} (physics), the distinctive requirement is a dipolar propagator from opposite dislocations. Codex and DeepSeek open the gold and confirm it. Claude searches for 48 calls, passes over the gold in one search result, settles on a recent model of deformed crystals, and concedes that the paper does not use that phrase. Kimi finds no arXiv match and names a 1994 activated model, crediting it with the gold's features. In Table~\ref{tab:qual-standard-stat} (stat), Codex and Claude open the gold and check the sparsity condition, while Kimi and DeepSeek note the gold in their search results, then follow a related method and never open it.

\paragraph{\RI: checking what must be inferred.}
In Table~\ref{tab:qual-implicit-qbio} (q-bio), whether a candidate's model has four parameters, is deterministic, and adds a linear final trend must be checked against its model equations. DeepSeek opens the gold and quotes the linear-trend equation. Codex returns a paper whose Richards curve has five parameters and calls its forecasts deterministic, although that paper models the counts as a stochastic process, and Kimi names a paper in which it never found the linear term. Claude runs out of time without opening the gold, which its searches return ten times. In Table~\ref{tab:qual-implicit-math} (math), the gold never says that it is non-Bayesian, ignores missing data, or uses only simulated data, and the sharp jump in its loss appears only in a figure. Codex and Claude open the gold and check each point, and Claude cites the figure. DeepSeek returns a paper that meets the other constraints but never tunes a regularization parameter and plots no loss curve, and Kimi reaches the message limit without an answer.

\paragraph{\CI: revising an earlier answer, or keeping it.}
In Table~\ref{tab:qual-cumulative} (eess), only the last turn reveals the constraints that single out the gold. Claude commits to RadioDUN at the third turn, then opens the gold at the fourth and explicitly drops its earlier answer. Kimi moves from one wrong paper to a title and id that do not match, and claims to have verified it. In Table~\ref{tab:qual-cumulative-econ}, the zero-count check at T3 points to GitHub activity data. Claude drops its T2 pick, an online experiment with no split by experience, and confirms the pattern in a figure of the gold. DeepSeek keeps the paper it found at T1 through all four turns and commits to it while saying it could not confirm the zero-count check, although one of its T2 search results named the gold's authors.

\begin{table*}[t]
\centering
\footnotesize
\renewcommand{\arraystretch}{1.15}
\begin{tabularx}{\textwidth}{@{}p{0.11\textwidth} p{0.2\textwidth} c c X@{}}
\toprule
\rowcolor{gray!12}\multicolumn{5}{@{}p{0.98\textwidth}@{}}{\textbf{Query} (\Std, physics).\ I need a study on plastic deformation in 2D crystalline and glassy materials that uses continuous modeling to track dynamical evolution on a periodic potential, since I'm looking for how collective effects manifest in 2D systems rather than threshold-based or activated descriptions. The work must specifically address the dipolar propagator arising from opposite dislocations, because I'm trying to understand how these long-range elastic interactions shape yielding in amorphous and crystalline solids without relying on stress thresholds or activated event durations.} \\
\rowcolor{gray!12}\multicolumn{5}{@{}p{0.98\textwidth}@{}}{\textbf{Gold.} \textit{Plastic Flow in Two-Dimensional Solids} (arXiv:cond-mat/0304209), a time-dependent Ginzburg--Landau model in which \evid{The elastic energy density is a periodic function of the shear and tetragonal strains} and whose slips are \evid{a pair of edge dislocations with opposite Burger vectors (dislocation dipoles)}.} \\
\midrule
Agent & Answer & Calls & Opened & Rationale (excerpt) \\
\midrule
\rowcolor{okGreen!10} Codex & Gold \textcolor{okGreen}{\ding{51}} & 6 & yes & \evid{its elementary slip consists of two edge dislocations with opposite Burgers vectors, producing the specified long-range dipolar elastic response.} \\
\rowcolor{badRed!8} Claude & \textit{Self-induced marginality in plastically deformed crystals} (2511.08187) \textcolor{badRed}{\ding{55}} & 48 & no & \evid{The one unverified constraint is the literal phrase `dipolar propagator,' which the paper does not use \dots, so confidence is tempered.} \\
\rowcolor{badRed!8} Kimi & Bulatov and Argon (1994, part III; not on arXiv)$^\dagger$ \textcolor{badRed}{\ding{55}} & 22 & no & \evid{a continuous local plastic variable evolves in a periodic potential, long-range coupling is represented by the elastic propagator of two opposite dislocations \dots} \\
\rowcolor{okGreen!10} DeepSeek & Gold \textcolor{okGreen}{\ding{51}} & 8 & yes & \evid{the elastic energy density being a periodic function of shear and tetragonal strains \dots, and with slips consisting of two edge dislocations of opposite Burgers vectors giving a dipolar propagator.} \\
\bottomrule
\end{tabularx}
\caption{\textbf{Four agents on one \Std query (physics).} The paper Claude returns states that \evid{the fixed linear elastic propagators are replaced by the solution `on the fly' of the corresponding nonlinear elasticity problems} and never mentions dislocation dipoles. A review of these models (arXiv:1708.09194) files the Bulatov--Argon series among models that \evid{set a yield rate based on a stress-biased Arrhenius law for thermal activation}, which the query rules out. $^\dagger$Forced final answer after the message limit.}
\label{tab:qual-standard}
\end{table*}

\begin{table*}[t]
\centering
\footnotesize
\renewcommand{\arraystretch}{1.15}
\begin{tabularx}{\textwidth}{@{}p{0.11\textwidth} p{0.2\textwidth} c c X@{}}
\toprule
\rowcolor{gray!12}\multicolumn{5}{@{}p{0.98\textwidth}@{}}{\textbf{Query} (\Std, stat).\ I'm working with sparse networks where degree heterogeneity is significant, and I need a spectral embedding approach that remains stable when the average degree grows just above a polylogarithmic threshold. The method should handle degree-corrected structures where community members lie on hyperplanes in latent space, and produce embeddings suitable for downstream clustering under these sparse conditions. I need the work that establishes asymptotic normality for rows of scaled eigenvectors. This approach should differ from the one in `Two-sample Test of Community Memberships of Weighted Stochastic Block Models' in both its sparsity condition and model specification, and it should also use a different sparsity condition than `Asymptotic Theory of Eigenvectors for Random Matrices with Diverging Spikes'. The pipeline should show how degree-corrected latent positions from the same community lie on a one-dimensional hyperplane, outputting a d-1-dimensional embedding that feeds into clustering algorithms like k-means or Gaussian mixture models.} \\
\rowcolor{gray!12}\multicolumn{5}{@{}p{0.98\textwidth}@{}}{\textbf{Gold.} \textit{Spectral clustering under degree heterogeneity: a case for the random walk Laplacian} (arXiv:2105.00987), which proves a central limit theorem for rows of the random-walk-Laplacian embedding, \evid{provided the sparsity factor satisfies $n\rho_n = \omega(\log^{4c} n)$}.} \\
\midrule
Agent & Answer & Calls & Opened & Rationale (excerpt) \\
\midrule
\rowcolor{okGreen!10} Codex & Gold \textcolor{okGreen}{\ding{51}} & 11 & yes & \evid{it proves asymptotically Gaussian row-wise errors for random-walk Laplacian spectral embeddings under a sparse generalized random dot-product graph \dots} \\
\rowcolor{okGreen!10} Claude & Gold \textcolor{okGreen}{\ding{51}} & 35 & yes & \evid{It establishes a central limit theorem (asymptotic normality) for the rows of the scaled eigenvectors of the random walk Laplacian, handles significant degree heterogeneity \dots} \\
\rowcolor{badRed!8} Kimi & ``The Two-Sample Testing Problem for Degree-Corrected Block Models'' (2112.09905)$^\dagger$ \textcolor{badRed}{\ding{55}} & 20 & no & \evid{This matches Joshua Agterberg's `The Two-Sample Testing Problem for Degree-Corrected Block Models' (arXiv:2112.09905) \dots} We found no paper with this title, and the id belongs to a laser-physics paper. \\
\rowcolor{badRed!8} DeepSeek & No answer$^\dagger$ \textcolor{badRed}{\ding{55}} & 20 & no & While searching: \evid{This is the key insight of SCORE. In SCORE, for K communities, the rows of the K leading eigenvectors of the adjacency matrix lie on K lines (1D hyperplanes) through the origin.} \\
\bottomrule
\end{tabularx}
\caption{\textbf{Four agents on one \Std query (stat).} Kimi and DeepSeek both note the gold in their search results, then follow SCORE, a related embedding whose communities also lie on lines, and never open it. Kimi's forced answer pairs a title we could not find with the id of an unrelated paper, and DeepSeek gives no answer. $^\dagger$Forced final answer after the message limit.}
\label{tab:qual-standard-stat}
\end{table*}

\begin{table*}[t]
\centering
\footnotesize
\renewcommand{\arraystretch}{1.15}
\begin{tabularx}{\textwidth}{@{}p{0.11\textwidth} p{0.2\textwidth} c c X@{}}
\toprule
\rowcolor{gray!12}\multicolumn{5}{@{}p{0.98\textwidth}@{}}{\textbf{Query} (\RI, q-bio).\ I'm modeling an infectious disease outbreak and need a method that captures the S-shaped trajectory as case counts rise and then plateau. The extra shape parameter is what lets the curve flex to fit different outbreak patterns, so I'm after a four-parameter sigmoidal model. I also want a deterministic formulation so I can get consistent forecasts without simulation noise, and I need the whole epidemic trajectory described by a single growth equation rather than a patchwork of phases. The method should include a linear final-trend adjustment for cases where the plateau isn't flat, like some countries' outbreaks. I'm not looking for a broad model comparison; I need the study that commits to a focused growth curve approach, employing fewer model types than `Rational evaluation of various epidemic models based on the COVID-19 data of China'.} \\
\rowcolor{gray!12}\multicolumn{5}{@{}p{0.98\textwidth}@{}}{\textbf{Gold.} \textit{Using generalized logistics regression to forecast population infected by Covid-19} (arXiv:2004.02406), which fits the curve $M/(1+e^{-at+b})^\alpha$ to cumulative cases and, for South Korea, adds a linear term, $(M+ct)/(1+e^{-at+b})^\alpha$.} \\
\midrule
Agent & Answer & Calls & Opened & Rationale (excerpt) \\
\midrule
\rowcolor{badRed!8} Codex & \textit{Nowcasting COVID-19 incidence indicators during the Italian first outbreak} (2010.12679) \textcolor{badRed}{\ding{55}} & 15 & no & \evid{The paper focuses on a Richards generalized logistic growth curve \dots\ supports deterministic point forecasts via maximum-likelihood estimates, and modifies the baseline with a linear temporal trend} \\
\rowcolor{badRed!8} Claude & No answer (time limit) \textcolor{badRed}{\ding{55}} & 57 & no & Runs out of time without an answer. The gold appears in its search results at ten steps, and it never opens it. \\
\rowcolor{badRed!8} Kimi & \textit{Generalized logistic growth modeling of the COVID-19 outbreak \dots} (2003.05681)$^\dagger$ \textcolor{badRed}{\ding{55}} & 20 & no & Before the limit: \evid{Let me try more specific phrase searches to verify the linear-trend component in the Wu et al.\ paper.} Forced answer: \evid{\dots\ with growth-model (linear-trend) treatment for countries whose cumulative case counts had not plateaued.} \\
\rowcolor{okGreen!10} DeepSeek & Gold \textcolor{okGreen}{\ding{51}} & 13 & yes & \evid{Critically, section 3.2 introduces equation (3.4): $P(t)=(M+ct)/(1+e^{-at+b})^\alpha$, a linear-trend-adjusted variant explicitly designed for outbreaks like South Korea where the plateau is not flat.} \\
\bottomrule
\end{tabularx}
\caption{\textbf{Four agents on one \RI query (q-bio).} The paper Codex returns uses a Richards curve that \evid{depends on 5 parameters}, and it treats each series as \evid{the realization of the stochastic process Z}, building its forecast intervals from simulated count series. The paper Kimi returns calibrates four growth models, none of which adds a linear trend to a plateau. $^\dagger$Forced final answer after the message limit.}
\label{tab:qual-implicit-qbio}
\end{table*}

\begin{table*}[t]
\centering
\footnotesize
\renewcommand{\arraystretch}{1.15}
\begin{tabularx}{\textwidth}{@{}p{0.11\textwidth} p{0.2\textwidth} c c X@{}}
\toprule
\rowcolor{gray!12}\multicolumn{5}{@{}p{0.98\textwidth}@{}}{\textbf{Query} (\RI, math).\ I'm trying to identify parametric spatiotemporal dynamical models where the coefficients exhibit step changes, but I'm struggling with the sharp transitions in the loss landscape when tuning regularization. I need a procedure that handles these coefficient variations without requiring Bayesian uncertainty quantification, and I want to avoid methods that get bogged down in handling missing data. I also need an approach that is illustrated with less real data than `DeepMoD: Deep learning for Model Discovery in noisy data', so I can focus on the core identification pipeline without that overhead.} \\
\rowcolor{gray!12}\multicolumn{5}{@{}p{0.98\textwidth}@{}}{\textbf{Gold.} \textit{Data-driven identification of parametric partial differential equations} (arXiv:1806.00732), which selects PDE terms by group-sparse regression, plots the loss over 50 threshold values (Fig.~2), and tests on four simulated PDEs, including a Navier-Stokes run whose Reynolds number \evid{is adjusted half way through the simulation}.} \\
\midrule
Agent & Answer & Calls & Opened & Rationale (excerpt) \\
\midrule
\rowcolor{okGreen!10} Codex & Gold \textcolor{okGreen}{\ding{51}} & 7 & yes & \evid{It analyzes model-selection loss while sweeping regularization thresholds, is deterministic rather than Bayesian, does not focus on missing-data recovery, and uses four synthetic canonical systems with no experimental dataset \dots} \\
\rowcolor{okGreen!10} Claude & Gold \textcolor{okGreen}{\ding{51}} & 18 & yes & \evid{Fig. 2 shows the loss exhibiting sharp step-like transitions as the ridge/threshold regularization ($\varepsilon$, $\lambda$) is tuned, the method (SGTR group sparse regression) is a deterministic point estimate \dots} \\
\rowcolor{badRed!8} Kimi & No answer$^\dagger$ \textcolor{badRed}{\ding{55}} & 25 & no & While searching: \evid{DeepMoD has one real experiment. `less real data than DeepMoD' then zero real datasets. Candidate likely synthetic only.} The gold appears once in its search results, and it never opens it. \\
\rowcolor{badRed!8} DeepSeek & \textit{Estimating Varying Parameters in Dynamical Systems \dots} (2412.16198) \textcolor{badRed}{\ding{55}} & 12 & no & \evid{fits piecewise-constant parameter-varying models using optimization-based parameter estimation (Nelder-Mead and Powell methods) rather than regularization-tuned sparse regression.} \\
\bottomrule
\end{tabularx}
\caption{\textbf{Four agents on one \RI query (math).} The paper DeepSeek returns fits switching parameters in simulated equations with a deterministic optimizer, so it meets the step-change, Bayesian, missing-data, and real-data constraints. It fails the loss-curve constraint: its only sparse regression fixes \evid{$\lambda = 0.01$}, and none of its figures plots loss against a regularization parameter. $^\dagger$Forced final answer after the message limit.}
\label{tab:qual-implicit-math}
\end{table*}

\begin{table*}[t]
\centering
\footnotesize
\renewcommand{\arraystretch}{1.15}
\begin{tabularx}{\textwidth}{@{}p{0.09\textwidth} p{0.13\textwidth} p{0.13\textwidth} p{0.2\textwidth} X@{}}
\toprule
\rowcolor{gray!12}\multicolumn{5}{@{}p{0.98\textwidth}@{}}{\textbf{Conversation} (\CI, eess; user turns abridged).
\textbf{T1.} I'm looking for a concrete scholarly method I can adapt for a learning-based engineering project with a tightly limited pipeline \dots
\textbf{T2.} \dots\ I want the physical assumptions encoded directly in the learned representation, not supplied as inputs or imposed through auxiliary optimization terms. \dots\ The way the method integrates physics must also differ from `TiRE-GAN: Task-Incentivized Generative Learning for Radiomap Estimation' \dots
\textbf{T3.} \dots\ the method should be tied to a specific architecture rather than being backbone-agnostic. Compared with `RadioGAT \dots', it needs to use different physical knowledge \dots
\textbf{T4.} \dots\ Transmitter-side paths should pair one-to-one with ellipsoids, while receiver-side rays are sampled independently of the ellipsoid count. Scattered paths must accumulate attenuation on both the transmitter and receiver sides, whereas the direct path does so only once; the results should also show faint multipath structure becoming visible after a logarithmic display transform \dots\ Please commit to one paper now \dots} \\
\rowcolor{gray!12}\multicolumn{5}{@{}p{0.98\textwidth}@{}}{\textbf{Gold.} \textit{6D Channel Knowledge Map Construction via Bidirectional Wireless Gaussian Splatting} (arXiv:2510.26166).} \\
\midrule
Agent & After T2 & After T3 & After T4 (final) & Excerpt \\
\midrule
\rowcolor{badRed!8} Codex & none & none & nGRF (2508.11668) \textcolor{badRed}{\ding{55}} & \evid{transmitter-to-Gaussian paths correspond to primitives while receiver rays are sampled separately; scattered components account for both path segments \dots} \\
\rowcolor{okGreen!10} Claude & RadioDUN (2506.08418) & RadioDUN & Gold \textcolor{okGreen}{\ding{51}} & At T3: \evid{So my committed answer is RadioDUN \dots} At T4, after opening the gold: \evid{This resolves it --- and it overturns my earlier RadioDUN pick, which was wrong for these mechanics.} \\
\rowcolor{badRed!8} Kimi & NeRF$^2$ (2305.06118) & NeRF$^2$ & ``WRF-GS: Planar \dots'' (2411.19356) \textcolor{badRed}{\ding{55}} & \evid{WRF-GS (2411.19356, full text and equations verified) satisfies all constraints \dots} The id belongs to an unrelated study of public perception of AI, and Kimi never read it. \\
\rowcolor{okGreen!10} DeepSeek & none & PeFNO / PERNN & Gold \textcolor{okGreen}{\ding{51}} & Names the gold from a search snippet without opening it. \\
\bottomrule
\end{tabularx}
\caption{\textbf{Four agents on one \CI conversation (eess).} Each column gives the paper an agent proposes after that turn. Codex's final paper computes the channel as a weighted sum over Gaussians, with interference carried by phases \evid{not through multiplicative attenuation}, and it samples no receiver rays, so it violates the T4 constraints on attenuation and ray sampling.}
\label{tab:qual-cumulative}
\end{table*}

\begin{table*}[t]
\centering
\footnotesize
\renewcommand{\arraystretch}{1.15}
\begin{tabularx}{\textwidth}{@{}p{0.09\textwidth} p{0.14\textwidth} p{0.12\textwidth} p{0.14\textwidth} X@{}}
\toprule
\rowcolor{gray!12}\multicolumn{5}{@{}p{0.98\textwidth}@{}}{\textbf{Conversation} (\CI, econ; user turns abridged).
\textbf{T1.} I'm trying to get rigorous evidence on whether changing access to a workplace aid alters performance disparities between people with different levels of experience. \dots
\textbf{T2.} For the less-experienced group, the paper has to show that removing AI access increased both novices' output quantity and their success rate. \dots
\textbf{T3.} \dots\ I need a distributional check because inactivity might be concentrated in one kind of work. The frequency of zero counts should be highest for issue-closing events and lowest when all event types are summed.
\textbf{T4.} To position this evidence alongside `Who is using AI to code? Global diffusion and impact of generative AI', the paper should use a different task design, analyze a smaller sample, and report more detailed quality outcomes. \dots\ please commit to one paper now \dots} \\
\rowcolor{gray!12}\multicolumn{5}{@{}p{0.98\textwidth}@{}}{\textbf{Gold.} \textit{The Heterogeneous Productivity Effects of Generative AI} (arXiv:2403.01964), which studies the GitHub activity of \evid{over 36,000 GitHub users} around Italy's ChatGPT ban.} \\
\midrule
Agent & After T2 & After T3 & After T4 (final) & Excerpt \\
\midrule
\rowcolor{okGreen!10} Codex & none & none & Gold \textcolor{okGreen}{\ding{51}} & At T3, without naming it: \evid{I've found a strong exact-match candidate. Its data show the required ordering of zero-count frequencies \dots} \\
\rowcolor{okGreen!10} Claude & \textit{AI Assistance Reduces Persistence \dots} (2604.04721) & Gold & Gold \textcolor{okGreen}{\ding{51}} & At T2: \evid{remove the AI and people attempt more and solve more.} At T3: \evid{This is your paper --- and I've now verified the distributional check directly against the figures \dots} \\
\rowcolor{badRed!8} Kimi & Otis et al.\ / Luo et al.\ (tentative) & Hoffmann et al.\ (not on arXiv) & Abstains \textcolor{badRed}{\ding{55}} & At T4: \evid{nothing I can locate and verify satisfies all of your binding constraints simultaneously.} \\
\rowcolor{badRed!8} DeepSeek & \textit{Generative AI at Work} (2304.11771) & Same & Same \textcolor{badRed}{\ding{55}} & At T4: \evid{Given your instruction to commit, I'm going with this paper despite being unable to 100\% confirm one distributional detail.} \\
\bottomrule
\end{tabularx}
\caption{\textbf{Four agents on one \CI conversation (econ).} Each column gives the paper an agent proposes after that turn. In DeepSeek's conversation, the simulated user wrongly confirmed at T3 that AI access was provided rather than removed, so we judge DeepSeek only on the zero-count check, which it received unchanged. \textit{Generative AI at Work}, which DeepSeek keeps from T1 to the end, uses \evid{data from 5,172 customer support agents} and has no GitHub event counts, so it cannot show that pattern.}
\label{tab:qual-cumulative-econ}
\end{table*}

\paragraph{\NM: finding the near-miss is not enough.}
Tables~\ref{tab:qual-unmet-physics} and~\ref{tab:qual-unmet-math} show two \NM queries. Each injected clause contradicts a detail that the near-miss reports, and all four agents open the near-miss in both cases. Codex abstains on both. In physics, Claude, Kimi, and DeepSeek each see that the near-miss maximizes J and still return it, calling the clause an inaccurate paraphrase, a misrecollection, or a sign flip. In math, Claude sees that the near-miss selects its equation with its own energy and guesses that a journal version uses AIC, which that version does not, and the forced answers of Kimi and DeepSeek state that the paper uses AIC, although the text each received was cut off before the paper's selection step.

\begin{table*}[t]
\centering
\footnotesize
\renewcommand{\arraystretch}{1.15}
\begin{tabularx}{\textwidth}{@{}p{0.11\textwidth} p{0.14\textwidth} c c X@{}}
\toprule
\rowcolor{gray!12}\multicolumn{5}{@{}p{0.98\textwidth}@{}}{\textbf{Query} (\NM, physics).\ I'm benchmarking a new generative model and need the most directly comparable prior work that uses recurrent networks to generate molecules from SMILES strings on the ZINC dataset. I specifically need the approach that trains on around 250,000 examples with a practical vocabulary of 64 tokens, since my own setup requires matching that scale to avoid distribution shifts. The method should employ an autoregressive generation strategy rather than variational or adversarial approaches, and I need to see how it integrates Monte Carlo Tree Search with RNN-based rollouts for the simulation step. \textcolor{colUnmet!80!black}{\textbf{For consistency with my benchmark, I need the reported molecules to be ranked by minimizing J, with the top molecule having the lowest objective score.}} This baseline must construct molecules through a search tree where children are sampled from the recurrent network multiple times during expansion, then completed via rollout to terminal nodes for reward computation.} \\
\rowcolor{gray!12}\multicolumn{5}{@{}p{0.98\textwidth}@{}}{\textbf{Gold.} $\varnothing$. \textbf{Near-miss.} \textit{ChemTS: an efficient python library for de novo molecular generation} (arXiv:1710.00616), in which \evid{effectiveness of each method is quantified by the maximum score J among all generated molecules}, and whose best molecules are listed by descending J.} \\
\midrule
Agent & Answer & Calls & Opened & Rationale (excerpt) \\
\midrule
\rowcolor{okGreen!10} Codex & Abstains \textcolor{okGreen}{\ding{51}} & 5 & yes & \evid{However, ChemTS explicitly maximizes J; its best molecule has the highest score (6.56), not the lowest. Therefore no paper satisfies all stated requirements.} \\
\rowcolor{badRed!8} Claude & Near-miss \textcolor{badRed}{\ding{55}} & 6 & yes & \evid{The lone mismatch \dots\ `minimizing J / lowest objective' vs.\ ChemTS maximizing J \dots\ is an inaccuracy in the paraphrase} \\
\rowcolor{badRed!8} Kimi & Near-miss \textcolor{badRed}{\ding{55}} & 7 & yes & \evid{the paper actually maximizes J (penalized logP-style reward; the top molecule has the highest J = 6.56), whereas the researcher recalled minimizing J} \\
\rowcolor{badRed!8} DeepSeek & Near-miss \textcolor{badRed}{\ding{55}} & 16 & yes & \evid{The only minor terminological discrepancy is ``minimizing J'' versus the paper's maximization framing, which is a trivial sign flip for benchmarking purposes.} \\
\bottomrule
\end{tabularx}
\caption{\textbf{Four agents on one \NM query (physics).} The clause asks for molecules ranked by minimizing the objective J, and the near-miss ranks them by the highest J. All four agents read the near-miss, and the three that return it dismiss the mismatch. None of the four is a forced final answer.}
\label{tab:qual-unmet-physics}
\end{table*}

\begin{table*}[t]
\centering
\footnotesize
\renewcommand{\arraystretch}{1.15}
\begin{tabularx}{\textwidth}{@{}p{0.11\textwidth} p{0.14\textwidth} c c X@{}}
\toprule
\rowcolor{gray!12}\multicolumn{5}{@{}p{0.98\textwidth}@{}}{\textbf{Query} (\NM, math).\ I'm trying to identify differential equation structures from frequency response data, but my current approach struggles when features have vastly different scales. I need the method that uses Fourier expansions with an l0 sparsity constraint and trims groups based on core-region energy to handle these scale variations without cross-validation failing in noisy settings. \textcolor{colUnmet!80!black}{\textbf{After core-energy group trimming has screened the terms, I need the final choice among surviving equations to rely solely on the Akaike information criterion.}} The approach should demonstrate this on the KdV equation to show it works in practice.} \\
\rowcolor{gray!12}\multicolumn{5}{@{}p{0.98\textwidth}@{}}{\textbf{Gold.} $\varnothing$. \textbf{Near-miss.} \textit{Fourier Features for Identifying Differential Equations (FourierIdent)} (arXiv:2311.16608), which finds the optimal sparsity \evid{by minimizing the energy E1 + E2} and cites Akaike only in its history of the field. Its journal version selects the same way.} \\
\midrule
Agent & Answer & Calls & Opened & Rationale (excerpt) \\
\midrule
\rowcolor{okGreen!10} Codex & Abstains \textcolor{okGreen}{\ding{51}} & 9 & yes & \evid{its surviving models are selected using a bespoke energy combining core-region fitting residual and coefficient stability, rather than solely the Akaike information criterion.} \\
\rowcolor{badRed!8} Claude & Near-miss \textcolor{badRed}{\ding{55}} & 20 & yes & \evid{The only detail I could not confirm on the arXiv v1 is that final selection uses solely the Akaike information criterion (v1 uses a residual+stability energy); this most likely reflects the substantially revised journal version \dots} \\
\rowcolor{badRed!8} Kimi & Near-miss$^\dagger$ \textcolor{badRed}{\ding{55}} & 18 & yes & \evid{\dots\ with a new core-region-based energy for coefficient identification, and selecting the final equation purely by AIC instead of cross-validation} \\
\rowcolor{badRed!8} DeepSeek & Near-miss$^\dagger$ \textcolor{badRed}{\ding{55}} & 19 & yes & \evid{The full text discusses model selection via AIC after group trimming, and the method is tested on the KdV equation among others. This paper is the clear match.} \\
\bottomrule
\end{tabularx}
\caption{\textbf{Four agents on one \NM query (math).} The clause asks that the final equation be chosen by the Akaike information criterion, and the near-miss chooses it by minimizing its own energy. Claude sees the mismatch and appeals to a journal version, which uses the same energy. The forced answers of Kimi and DeepSeek claim that the paper uses AIC, although the text each received was cut off before the selection step; DeepSeek's claim follows a search snippet of the journal version's reference list, which cites Akaike. $^\dagger$Forced final answer after the message limit.}
\label{tab:qual-unmet-math}
\end{table*}

\end{document}